\documentclass[10pt,conference]{IEEEtran}
\IEEEoverridecommandlockouts

\usepackage{amsmath,amssymb,amsfonts}
\usepackage{graphicx}
\usepackage{subcaption}
\usepackage[table,xcdraw]{xcolor}
\usepackage{balance}
\usepackage{soul}
\usepackage{hyperref}
\usepackage{multirow}
\usepackage{fancybox}
\usepackage{enumitem}
\usepackage{graphicx}
\usepackage{color}
\usepackage{float}
\usepackage{xcolor}
\usepackage{array}
\usepackage{amsmath}
\usepackage{url}
\usepackage{tikz}
\usepackage{pgfplots}
\usepackage{listings}
\usepackage{color}
\usepackage{graphicx}
\definecolor{dkgreen}{rgb}{0,0.6,0}
\definecolor{gray}{rgb}{0.5,0.5,0.5}
\definecolor{mauve}{rgb}{0.58,0,0.82}
\pgfplotsset{compat=1.16}
\usepackage{pgf-pie}
\usetikzlibrary{pgfplots.statistics,calc}
\usepackage{algorithm}
\usepackage{algpseudocode}
\usepackage{listings}

\usepackage{tcolorbox}

\makeatletter
\newcommand{\mybox}[1]{%
	\setbox0=\hbox{#1}%
	\setlength{\@tempdima}{\dimexpr\wd0+13pt}%
	\begin{tcolorbox}[boxrule=0.5pt, colback=gray!10, arc=4pt,
		left=6pt,right=6pt,top=6pt,bottom=6pt,boxsep=0pt]
		#1
	\end{tcolorbox}
}

\definecolor{codegreen}{rgb}{0,0.6,0}
\definecolor{codegray}{rgb}{0.5,0.5,0.5}
\definecolor{codepurple}{rgb}{0.58,0,0.82}
\definecolor{backcolour}{rgb}{0.95,0.95,0.92}

\newcommand{\tikzcircle}[2][red,fill=red]{\tikz[baseline=-0.5ex]\draw[#1,radius=#2] (0,0) circle ;}%

\lstdefinestyle{mystyle}{
  language=Python,
  aboveskip=3mm,
  showstringspaces=false,
  columns=flexible,
  numbers=none,
  backgroundcolor=\color{backcolour},
  commentstyle=\color{codegreen},
 keywordstyle=\color{magenta},
    numberstyle=\tiny\color{codegray},
    stringstyle=\color{codepurple},
    basicstyle=\small\ttfamily,
    breakatwhitespace=false,         
    breaklines=false,                 
    captionpos=b,                    
    keepspaces=false,                 
    numbersep=5pt,                  
    showspaces=false,                
    showstringspaces=false,
    showtabs=false,                  
    tabsize=2,
    escapeinside=``
}
\def\BibTeX{{\rm B\kern-.05em{\sc i\kern-.025em b}\kern-.08em
    T\kern-.1667em\lower.7ex\hbox{E}\kern-.125emX}}
\usepackage{booktabs} 

\usepackage{tabularx}
\usepackage{pgfplots}

\begin{document}

\title{A First Look at Fairness of Machine Learning Based Code Reviewer Recommendation}

\author{
    \IEEEauthorblockN{
     Mohammad Mahdi Mohajer\IEEEauthorrefmark{1}, 
     Alvine Boaye Belle\IEEEauthorrefmark{1},
     Nima Shiri harzevili\IEEEauthorrefmark{1},
    Junjie Wang\IEEEauthorrefmark{2},\\
    Hadi Hemmati\IEEEauthorrefmark{1}, 
    Song Wang\IEEEauthorrefmark{1},
   Zhen Ming (Jack) Jiang\IEEEauthorrefmark{1}}
    \IEEEauthorblockA{\IEEEauthorrefmark{1}York University; \IEEEauthorrefmark{2}Institute of Software, Chinese Academy of Sciences;
     \\\{mmm98,alvine.belle,nshiri,hemmati,wangsong,zmjiang\}@yorku.ca; junjie@iscas.ac.cn}
}

\maketitle

\begin{abstract}
	The fairness of machine learning (ML) approaches is critical to the reliability of modern artificial intelligence systems. Despite extensive study on this topic, the fairness of ML models in software engineering (SE) domain has not been well explored yet. 
As a result, many ML-powered software systems, particularly those utilized in software engineering community, continue to be prone to fairness issues.  {Taking one of the typical SE tasks, i.e., code reviewer recommendation, as a subject, this paper conducts the first study toward investigating the issue of fairness of ML applications in the SE domain.} Our empirical study demonstrates that current state-of-the-art ML-based code reviewer recommendation techniques exhibit unfairness and discriminating behaviors. {Specifically, male reviewers get on average 7.25\% more recommendations than female code reviewers compared to their distribution in the reviewer set.} This paper also discusses the reasons why the studied ML-based code reviewer recommendation systems are unfair and provides solutions to mitigate the unfairness. 
Our study further indicates that the existing mitigation methods can enhance fairness by 100\% in projects with a similar distribution of protected and privileged groups, but their effectiveness in improving fairness on imbalanced or skewed data is limited. Eventually, we suggest a solution to overcome the drawbacks of existing mitigation techniques and tackle bias in datasets that are imbalanced or skewed.


\end{abstract}

\begin{IEEEkeywords}
Fairness, machine learning, reviewer recommendation
\end{IEEEkeywords}

\section{Introduction}
\label{sec:intro}

Machine Learning (ML) approaches and models are increasingly being used in the development of modern software~\cite{survey-deep4se} to assist developers in different tasks, e.g., defect prediction~\cite{defect-prediction1, defect-prediction2}, software bug triage~\cite{bug-triage1}, and code reviewer recommendation~\cite{WhoReview, corms}, etc. 
Meanwhile, the wide adoption of ML has given rise to new concerns and issues regarding the trustworthiness and ethicality of such systems, one of which is the issue of fairness \cite{Pessach2023, bias-fairness-survey2021}. In machine learning, fairness means ensuring that algorithms and models do not favor or discriminate against specific groups or people based on their human-related and demographic aspects such as race, gender, age, or ethnicity \cite{making-fair-ml,bias-fairness-survey2021,Pessach2023, fairness-enhancing-ml-interventions}.
Existing studies have shown that unfair machine learning applications can have serious impacts on people's lives~\cite{bias-discrimination-society, discrimination-aware-dm, homework-raising-awareness-fairness, fairness-criminal-justice, fairness-credit-scoring}. For instance, Amazon's ML-based hiring algorithm discriminates against female candidates~\cite{Pessach2023}. The fairness problem has also been shown to exist in other human-sensitive areas using machine learning for making decisions, such as judicial evaluation systems, credit scoring, and loan application filtering \cite{Pessach2023, bias-discrimination-society, fairness-criminal-justice, fairness-credit-scoring}.

Despite the fact that previous studies \cite{ltdd,fair-preprocessing,fairea,fair-smote,fairway,fairness-crowd,fairness-aware-cfg,ignorance-se-fairness,software-fairness-fse18, software-fairness-survey} have extensively examined the fairness of ML applications, while most of these studies mainly focus on general ML applications, 
little is known about the fairness of ML applications in software engineering domain, e.g., automated bug triage \cite{bug-triage1}. 
In this work, we take one of the typical SE tasks, i.e., code reviewer recommendation as a subject to explore the fairness of ML applications in SE domain. 
Specifically, code reviewer recommendation systems are widely used in modern software development to identify the most appropriate code reviewers for a code change.
Recently, many ML-based code reviewer recommendation systems have been proposed. For instance, Patanamon et al. proposed RevFinder that used a similarity of previously reviewed file path to recommend an appropriate code-reviewer~\cite{revfinder} and Pandya et al.~\cite{corms} proposed CORMS, which leveraged similarity analysis and support vector machine (SVM) models to recommend reviewers. Although these examined code reviewer recommendation approaches can achieve good performance, none of the fairness characteristics (e.g., {race}, {age}, and gender) were considered when recommending reviewers. As a result, there may be fairness issues in such systems that have not previously been investigated, which can potentially have an adverse impact on reviewers' activities.~\cite{software-fairness-fse18, Pessach2023, software-fairness-survey}.

To address the aforementioned concerns, in this paper we look into the problem of assessing fairness issues in ML-based code reviewer recommendation systems. To the best of our knowledge, this is the \textbf{first step} toward an empirical study that addresses the fairness problem in such systems in software engineering domain.
Specifically, 
we conduct an empirical study on two recent ML-based code reviewer recommendation systems, RevFinder~\cite{revfinder} and CORMS~\cite{corms} and 
we used the same dataset from CORMS~\cite{corms} to build these systems and run our experiments. 
{Note that, when exploring the fairness of ML-based code reviewer recommendation systems, we only consider the factor of gender. This is mainly because collecting data to identify sensitive factors such as age or race is difficult, as reviewers and code review platforms often do not disclose this information. Gender information is also often not specified by the reviewers. However, compared to other factors, this information can still be derived from other sources, e.g., social media accounts, personal websites, institutional websites, and GitHub profiles (details are provided in Section~\ref{sec:dc}).} 
Our experimental results show that both RevFinder and CORMS have unfair behavior in their recommendations. Specifically, they favor male reviewers over female reviewers. 
For example, we observe that in the Node.js project, male reviewers recommended by CORMS have approximately 85\% more chance of being recommended for new code review requests, which is 16\% more than the fair condition (details are in Section \ref{sec:result}). 
We further explore the underlying factors that contribute to the unfairness of ML-based code reviewer recommendation systems, e.g., popularity bias, and whether the existing unfairness mitigation approaches~\cite{fa-ir,detgreedy,fairness-recsys-preproc2, fairness-recsys-preproc1} can help improve the fairness of these systems. 
{Our experiment results show that the existing mitigation approaches can improve fairness, but not consistently across all projects. Thus, we further propose a new mitigation method to counteract the limitations of existing unfairness mitigation strategies. Our proposed technique has the potential to enhance fairness in multiple circumstances in code reviewer recommendation systems.} 
As a summary, this paper makes the following contributions:

\begin{itemize}
    \item We conduct an empirical study to investigate the fairness of two state-of-the-art code reviewer recommendation systems. To the best of our knowledge, this is the first study on fairness analysis of ML applications in software engineering domain.
    \item We analyze the underlying factors that influence the outcomes of code reviewer recommendation systems and demonstrate that the existing unfairness mitigation methods can be utilized to alleviate the unfairness in ML-based code reviewer recommendation systems. The improvement can be as high as 100\%, effectively removing bias from recommendations, particularly in projects where protected and privileged groups have a balanced distribution.
    \item We show that the current mitigation approaches have limitations in terms of fairness improvement for projects with imbalanced or skewed data. Motivated by that, we propose a new unfairness mitigation algorithm to solve the issues associated with the previous techniques. Our algorithm can significantly outperform the two examined baselines.
    \item We release the dataset and source code of our experiments to help other researchers replicate and extend our study\footnote{\url{https://doi.org/10.5281/zenodo.7897538}}. 
\end{itemize}

The rest of this paper is organized as follows. 
Section~\ref{sec:motivation} presents
the background and related studies of this work.  
Section~\ref{sec:approach} and Section~\ref{sec:result} show the experimental setup and the evaluation results, respectively. 
Section~\ref{sec:discussion} explores new approach to improve the fairness mitigation of code reviewer recommendation systems. Section~\ref{sec:threats} discusses the threats to validity of this work.
Finally, Section~\ref{sec:conclusion} concludes this paper.

\section{Background and Related Work}
\label{sec:motivation}



\subsection{Code Reviewer Recommendation System}
\label{sec:crrs}
A code reviewer recommendation system is a software application that supports software development teams in identifying the most appropriate code reviewers for a particular code change request by suggesting a list of the most qualified candidates to conduct the review request~\cite{mcr-survey2021}. 





While the first code reviewer recommendation system already utilized machine learning techniques when it was introduced \cite{first-crr}, earlier systems often relied on heuristic approaches \cite{revrec, rstrace}, such as graph and search-based approaches. As an example, Ounti et al. \cite{revrec} proposed RevRec, a recommendation system that uses a genetic algorithm to find an appropriate peer reviewer for a code change.
As the field progressed, newer systems increasingly employed more machine learning methods \cite{hybrid-colab, context-aware-crr, crr-expand, corms} e.g., SVM, collaborative filtering, and Naive Bayes, to improve their recommendations. 
These reviewer recommendation systems use different factors and features for determining the most qualified reviewer for a review request, such as file similarity, developers' expertise, social relations, developers' activeness, etc~\cite{mcr-survey2021}.

In this work, we select two state-of-the-art ML-based code reviewer recommendation systems, i.e., RevFinder~\cite{revfinder} and CORMS~\cite{corms}, as our research subjects to explore the fairness (details  are in Section \ref{sec:subs}).
\subsection{Fairness Analysis in Machine Learning Application}
\label{sec:fairness-ml}


\subsubsection{Fairness}
\label{sec:fairness}
In ML, fairness analysis is the process of figuring out if the algorithms, models, or systems that are being built are fair and treat everyone the same way, no matter their human-related or demographic information such as race, gender, or age. The aim is to ensure that the systems do not reinforce existing biases or discrimination and to prevent negative impacts on individuals or groups due to these biases~\cite{Pessach2023, bias-fairness-survey2021, how-do-fairness-definitions-fare, trustworthyai23}. 
In general, there are two major  causes of unfairness in machine learning applications, i.e., data bias, e.g., a dataset contains skewed judgments, reports, and measures, an imbalanced attribute distribution) and algorithm bias, e.g., the learning process may become biased and unfair if the model is optimized for performance at the expense of other factors. 

\subsubsection{Types of Fairness}
\label{sec:tof}
In general machine learning applications, there are two types of fairness~\cite{bias-fairness-survey2021, Pessach2023}:

\begin{itemize}
    \item Group Fairness: This type of fairness ensures that different groups of people, including protected groups and privileged groups, are treated similarly and fairly. For example, in cases where gender is not a deciding factor, the female group should be treated similarly to the male group~\cite{bias-fairness-survey2021, Pessach2023}.
    \item Individual Fairness: This type of fairness ensures that individuals that are similar based on a criterion should be treated fairly and similarly~\cite{bias-fairness-survey2021, Pessach2023, how-do-fairness-definitions-fare}. For example, regardless of demographic background, each applicant in the employment process should be treated equally.
\end{itemize}

{In this study, we conduct our analysis based on the group fairness definition. Rather than focusing on individual cases, group fairness analysis evaluates the influence of machine learning models on distinct groups of people. This means that group fairness analysis can assist in identifying and mitigating biases that may not be visible at the individual level but have a major influence on specific groups of people.}

\subsubsection{Unfairness in Recommendation Systems}
\label{sec:unfairness-recsys}
In this work, we focus on the fairness of recommendation systems. 
The concepts and definitions of fairness analysis in recommendation systems and general ML applications slightly differ~\cite{fairness-recsys-survey}. For example, in previous studies~\cite{Pessach2023, ltdd, bias-fairness-survey2021}, since all approaches target classification problems, the concept of fairness thoroughly depends on the predictions of the target attribute and its relation to the protected group. On the contrary, the concept of fairness in recommendation systems can be discussed from several points of view, e.g., the fairness of each of the recommended items (item-based fairness) and the fairness of the exposure and quality that each user experiences from the recommended items (user-based fairness) \cite{fairness-recsys-survey}. Also, in recommendation systems, not all of the biases are considered as unfairness, e.g., popularity bias, position bias, and conformity bias~\cite{fairness-recsys-survey, bias-recsys-survey}. The fairness of recommendation systems can be categorized into two groups \cite{fairness-recsys-survey}:

\begin{itemize}
    \item Process fairness: This means ensuring that the process used to produce recommendations is fair and unbiased to all users, regardless of their sensitive attributes.
    \item Outcome fairness: This means ensuring that the system's outcomes are distributed fairly and proportionately among various groups of people. This means that recommendations should neither favor nor discriminate against any particular group based on their sensitive attributes.
\end{itemize}



In this study, we focus on investigating \textbf{outcome fairness} in ML-powered code reviewer recommendation systems, specifically we examine the {item-based fairness} through \textbf{group fairness} analysis. In particular, the items being recommended in our case are code reviewers. Therefore, our research focuses on assessing whether the final list of recommendations upholds fairness with regards to these recommended code reviewers.





\subsection{Unfairness Mitigation Techniques}
\label{sec:ma}
To mitigate the unfairness in ML applications, many unfairness mitigation mechanisms have been proposed~\cite{bias-fairness-survey2021, Pessach2023}, which can be categorized into three major types.


\subsubsection{Pre-Processing Techniques}
\label{sec:preproc}
Pre-processing strategies \cite{fairness-recsys-preproc1, fairness-recsys-preproc2} aim to modify training data before it is utilized for training. The data itself is a common source of prejudice or unfairness. We may alleviate this unfairness even before the model attempts to learn it by applying pre-processing approaches.

\subsubsection{In-Processing Techniques}
\label{sec:inproc}
In-processing approaches \cite{fairness-recsys-inproc1, fairness-recsys-inproc2} alter the machine-learning algorithm to increase fairness. For example, regularization can be used as one of the in-processing strategies to increase fairness in machine learning models~\cite{fairness-recsys-survey, Pessach2023}. In-processing approaches can help decrease discrimination and ensure that the model delivers fair results for all groups by introducing fairness constraints into the model training process. Otherwise, the model will be penalized for unfairness. 

\subsubsection{Post-Processing Techniques}
\label{sec:postproc}
In a post-processing technique \cite{fairness-recsys-reranking1, fairness-recsys-reranking2}, the output scores and predictions of the machine learning model are processed after the learning process has taken place in order to make them more fair.

In this work, we focus on the ``post-processing'' category of fairness mitigation methods and exclude both ``in-processing'' and ``pre-processing'' mitigation approaches, as these two approaches either require non-trivial changes to the code reviewer recommendation systems, which are not generalizable, or are not suitable for the recommendation tasks used in our study subjects, thus not suitable for fairness study in recommendation systems~\cite{fairness-recsys-preproc1, fairness-recsys-preproc2}.



\subsection{Fairness in Software Engineering Domain}
\label{sec:se-fairness}

Many studies have been done on addressing the problem of fairness in general ML-based applications~\cite{software-fairness-fse18, fairsquare, fairness-mcr}, while little is known about the fairness of ML applications in software engineering domain, leaving the software systems vulnerable to fairness issues and discrimination~\cite{software-fairness-fse18, software-fairness-survey}.
Brun and Meliou \cite{software-fairness-fse18} outlined a vision for how software engineering research may address fairness flaws and called on the software engineering research community to act. They argued that software fairness is identical with software quality and that different software engineering difficulties, such as requirements, specification, design, testing, and verification, must be addressed in order to tackle this problem. 
Albarghouthi et al.~\cite{fairsquare} simplified the problem to a series of weighted volume computation problems and solved them using an SMT solver. They further developed a tool, i.e., FairSquare, to validate the fairness qualities of decision-making programs derived from real-world datasets. 
Some other research studies~\cite{fairness-mcr,crr-labeling-bias} work on the issue of bias in code reviewer recommendation in the process of modern code review. German et al.~\cite{fairness-mcr} investigated whether modern code reviews are conducted fairly, using fairness theory to develop a framework for understanding how fairness affects code reviews. Tecimer et al.~\cite{crr-labeling-bias} worked on fixing bias caused by labeling errors in order to improve the performance of code reviewer recommendation. Unlike these studies, we do not focus on the process of modern code review or the biases that are not related to fairness issues.
Our work is the first to look at how automatic code reviewer recommendation systems systematically discriminate against particular protected groups and analyze fairness based on human-related qualities.

\section{Empirical Study Setup}
\label{sec:approach}


For our analysis, we evaluate whether the recommendations made for a specific group of reviewers (i.e., male or female) are fair and, if not, how we can mitigate the issue. This section describes our research questions (Section~\ref{sec:rqs}), study subjects (Section~\ref{sec:subs}), experiment configurations (Section~\ref{sec:cfg}), data collection (Section~\ref{sec:dc}), evaluation measures (Section~\ref{sec:em}), and the details of our selected unfairness mitigation strategies (Section~\ref{sec:sfa}).

\subsection{Research Questions}
\label{sec:rqs}
In this study, we are going to answer the following three research questions (RQs):

\textbf{RQ1 (Existence of Fairness): Is there any unfairness in the ML-based code reviewer recommendation systems?}

This RQ analyzes fairness in two typical ML-based code reviewer recommendation systems using benchmark data from four open-source software repositories (details are in Section~\ref{sec:dc}). The findings of this section demonstrate the occurrence of discriminatory behavior and unfairness in code review systems in SE domain. 

\vspace{4pt}
\textbf{RQ2 (Root Cause for Unfairness): What is the root cause for unfairness in the code reviewer recommendation systems?}

In this RQ, we aim to investigate the factors that lead to unfairness in the results of recommendations in the studied subjects. These factors can come from several sources, such as an imbalanced dataset and popularity bias (details are in \ref{sec:answer_rq2}).

\vspace{4pt}
\textbf{RQ3 (Effectiveness of Existing Unfairness Mitigation Techniques): Do existing unfairness mitigation approaches work for ML-based code reviewer recommendation systems?}

In this RQ, we examine the effectiveness of existing unfairness mitigation solutions in code review recommendation systems. Specifically, we apply current fairness-improving strategies in the literature to the two studied ML-based code reviewer recommendation systems and further check whether they can help mitigate the unfairness problems.

\subsection{Subjects of Study}
\label{sec:subs}





{As mentioned in Section~\ref{sec:crrs}}, we use two state-of-the-art ML-based code reviewer recommendation systems as the research subjects, i.e., CORMS~\cite{corms} and RevFinder~\cite{revfinder}. RevFinder recommends code reviewers based on a machine learning-based ranking algorithm that learns the scores of different candidates for a given review request based on the similarities in the file locations involved in past code reviews from the dataset. By evaluating the records in the dataset and propagating the scores of each reviewer involved in review requests with similar file path locations, RevFinder provides a list of unique reviewers and their scores for the given dataset. On the other hand, CORMS 
uses the same similarity model as RevFinder (with additional features) and also employs a SVM model that learns from each review subject in the training set. Both systems apply combination techniques to combine the results into a single score list. 

Although these two code reviewer recommendation approaches that we have examined can achieve good performance in terms of accuracy of the reviewer recommendation task, none of the fairness characteristics, e.g., {race}, {age}, and gender, were considered when recommending reviewers by using these approaches.

\subsection{Configurations for Fairness Analysis}
\label{sec:cfg}

This section describes the configurations we use to conduct our fairness analysis experiments.

Our work focuses only on a demographic and human-related fairness factor, i.e., gender, as the attribute of interest for fairness analysis. 
Following previous fairness analysis studies~\cite{ltdd, Pessach2023, fairway, fair-smote, bias-fairness-survey2021}, we limit our analysis to binary gender values of male and female for simplicity's sake and because of the difficulty of collecting non-binary gender data. In this study, the female reviewer group is considered the protected group. Furthermore, we ensure the validity of our findings by replicating the configurations of the selected two code reviewer recommendation systems as specified in their respective publications~\cite{revfinder, corms}. The results of replication are confirmed by comparing them to the results of the papers, and the authors of these works ensure an accurate presentation of their work. When training each model, following existing studies~\cite{revfinder, corms}, we divide the dataset into two sections, with 80\% for training and the remaining 20\% for testing, chronologically. 
 In this work, we run CORMS and RevFinder on a PC equipped with a 2.8GHz i7-7700HQ processor and 16GB of RAM.

\begin{figure*}[!t]
\begin{center}

    \begin{tikzpicture}
    \begin{axis}[
        ybar,
        bar width=8pt,    
        width=\textwidth,    
        height=4cm,    
        ylabel={},    
        xlabel={},    
        xtick={1,2,3,4,5,6,7,8,9,10,11,12,13,14,15,16,17,18,19,20,21,22,23,24,25,26,27,28,29,30,31,32,33,34},
        xticklabel style={rotate=90},
        xticklabels={\textbf{nodejs}, openstack, \textbf{bssw}, unlegacy, oranse, opendaylight, pixel, \textbf{joyent}, eclipse, openbmc, android, \textbf{nixcommunity}, software factory, FDio, chromium, h5bp, onap, \textbf{getsentry}, libreoffice, fullstorydev, \textbf{shopify}, facebook, qt, \textbf{freeCodeCamp}, tensorflow, renovatebot, gerrit, opencord, mano, twbs, go, EbookFoundation, lineageOS, cloudera},    
        ymin=0,    
        ymax=70,
        xmin=0,
        xmax=35,
        title={Missing Names and Genders},
        every axis plot/.append style={
          ybar,
          bar shift=0pt,
          fill
        }
    ]

     \addplot[fill=red,fill opacity=0.5] coordinates {
        (1, 2.7)
        (3, 5.2)
        (8, 0)
        (12, 7.1)
        (18, 4.1)
        (21, 2.8)
        (24, 0)
    };
    
    \addplot[fill=blue,fill opacity=0.5] coordinates {
        (2, 36.1)
        (4, 68.7)
        (5, 60.8)
        (6, 19)
        (7, 55.8)
        (9, 37)
        (10, 64.9)
        (11, 63)
        (13, 16.3)
        (14, 13.3)
        (15, 62.1)
        (16, 12.5)
        (17, 21.1)
        (19, 64)
        (20, 13.7)
        (22, 14.2)
        (23, 60.5)
        (25, 14.9)
        (26, 16.6)
        (27, 51.6)
        (28, 53.4)
        (29, 68.5)
        (30, 23)
        (31, 34.4)
        (32, 25)
        (33, 32.8)
        (34, 13)
    };
    
    \addplot[const plot, dashed, red, line width=2pt] coordinates {(0,10) (35,10)};
    
    \end{axis}
    \end{tikzpicture}
\end{center}

\caption{The percentages of missing names and genders (``unknown'') for each project examined in our study. 
We select projects with at most 10\% missing names and genders, which is any project whose bar is beneath the red dashed line.} 
\label{fig:missing-genders}
\end{figure*}
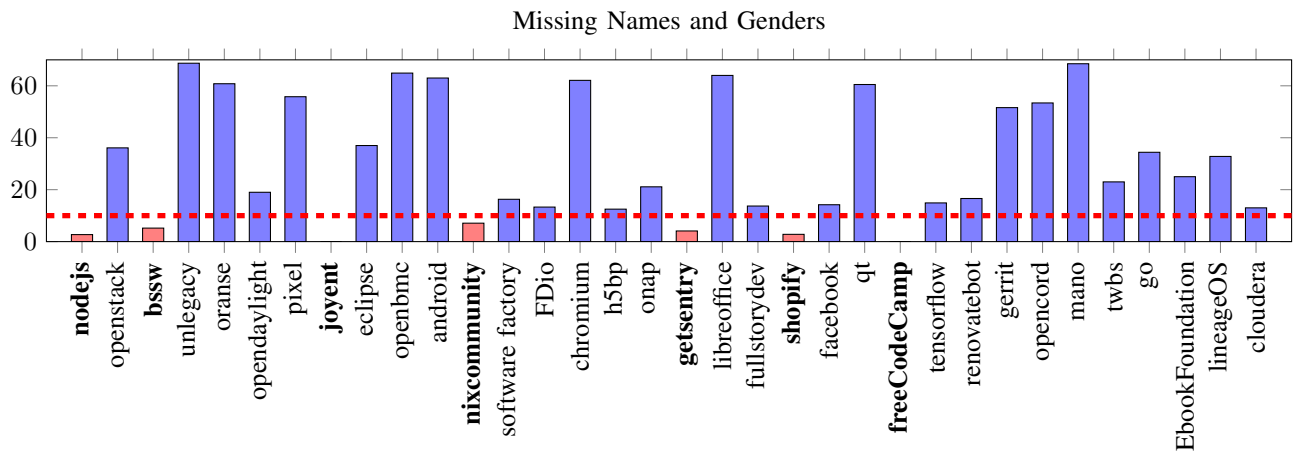

\subsection{Data Collection}
\label{sec:dc}

For our analysis, we need datasets that include human-related information of reviewers, such as their gender, which is our desired sensitive attribute. 
However, existing code reviewer recommendation datasets from previous studies~\cite{corms, revfinder, fairness-recsys-survey} do not include that critical information. 
{To collect the gender information of reviewers, in this work, we propose 
heuristic methods to infer a reviewer's gender information from their names and other publicly available information online (e.g., homepage, GitHub account information, and LinkedIn profile).}
Our experiment is conducted on the same dataset as the previous study \cite{corms}, which includes code review requests from 34 open-source projects. These code review requests are collected from code review platforms like Gerrit and GitHub that are available to the public. 
As mentioned before, these code review systems do not keep reviewers' human-related information, such as gender, and we cannot directly obtain the genders of the reviewers from the datasets.  

For each project in our dataset, we first get reviewers' names through the user ID provided by the code review platform. Then we use the following steps to infer a reviewer's demographic gender. 



\begin{enumerate}

    \item We first remove projects that have a high percentage of reviewers with missing names. If the field for a reviewer's name is null or blank, we consider this as not having a name mainly because the reviewer did not specify a name on his/her profile. Also, reviewers may have nicknames instead of their real names on their profiles. We consider all these records ``unknown'' since we cannot infer the gender from them. To distinguish nicknames from real names, we use the following heuristic, i.e., if a name contains any number, symbol, or sign, we consider that a nickname; e.g., in the nixcommunity project, there was a reviewer with the name ``jD91mZM2'', which has been removed from the data. We discard projects with reviewers who have more than 10\% ``unknown'' names to ensure the quality of our experiment data. The missing rates are demonstrated in Figure~\ref{fig:missing-genders}. As a result, we selected seven out of 34 projects. In the seven projects 
    examined, there were at most four reviewers who had nicknames on their profiles.
    \item Second, for the selected seven projects, we manually check each reviewer's information through online resources, e.g., personal and institutional homepages, GitHub accounts, and LinkedIn profiles. Some of the reviewers have links to their social media accounts on their GitHub profiles, so we can access their information directly. For the reviewers who do not provide social connections (20\% of our reviewer set), we search their full name on the internet to find their social media accounts and get the gender information. During the process, we discard names that point to indistinguishable users (e.g., people who have the same name and similar profiles on social media). 
    
    We look for gender information in their specified pronouns in their profiles, their online content, including the pronouns and genders the reviewers themselves or others used to describe them. 
    In this manual analysis, we checked 355 reviewers' information on the internet. According to our analysis, we obtained at least 90\% of the reviewers' gender information through this manual analysis. Multiple authors are involved during this process to ensure the correctness of the manually analyzed results.
    
    \item Third, those reviewers for whom we were unable to obtain gender information through manual analysis, we use Genderize.IO~\cite{genderize}, a third-party statistical gender prediction API. This tool takes a person's first name as an input parameter and uses statistical methods to guess the person's gender. Using this tool, we assigned gender information to six reviewers for whom we were not able to get their information through the manual analysis. It should be noted that there are reviewers with gender-neutral names, and Genderize.IO does not work on them. Therefore, if we cannot determine their gender through manual analysis, we exclude these reviewers from our dataset. Out of the selected seven projects, 25 reviewers had gender-neutral names. We were able to determine the gender of 23 of them, and we removed two reviewers who could not be identified.

    
    

    \item Finally, we remove projects for which there is only one reviewer from the protected group, i.e., the female reviewer group. In each remaining project, we exclude records relating to reviewers of unknown genders.
\end{enumerate}

After applying our gender identification approach, four out of the 34 candidate projects from~\cite{corms} are selected for our experiments, as demonstrated in Table~\ref{table:datasets}. 
Our strategy ensures the reliability and validity of our results by guaranteeing that each of the projects we use contains at least 90\% of its total reviewers and multiple people from different gender groups.

\begin{table}[]
\caption{Details of the four selected projects. The columns ``Missing Genders'', ``Total Ex.'', ``F. Ratio'', ``M\#'', and ``F\#'', are the percentage of missing genders, the total number of examples, the ratio of female reviewers, the number of female reviewers, and the number of male reviewers, respectively.}
\centering
\begin{tabular}{|c|c|c|c|c|c|}
\hline
Project               & Missing Genders & Total Ex.     & F.Ratio      & F\#      & M\#       \\ \hline
\textbf{nodejs}       & \textbf{2.7\%}  & \textbf{110} & \textbf{0.13} & \textbf{5}  & \textbf{28}  \\ \hline
\textbf{bssw}         & \textbf{5.2\%}  & \textbf{213} & \textbf{0.67} & \textbf{9}  & \textbf{9}  \\ \hline
\textbf{getsentry}    & \textbf{4.1\%}  & \textbf{775} & \textbf{0.08} & \textbf{6}  & \textbf{64}  \\ \hline
\textbf{shopify}      & \textbf{2.8\%}  & \textbf{565} & \textbf{0.06} & \textbf{17} & \textbf{148} \\ \hline
\end{tabular}
\label{table:datasets}
\end{table}

\subsection{Evaluation Measures}
\label{sec:em}

This section describes our evaluation measures for fairness and recommendation performance. 
Existing studies~\cite{fairness-acc-tradeoff2, fairness-enhancing-ml-interventions} revealed that improving fairness usually comes at a cost in terms of performance measures such as model accuracy. As a result, in order to demonstrate that an unfairness mitigation strategy is useful to deploy, most studies in fairness analysis will include a performance measure before and after applying the unfairness mitigation technique~\cite{Pessach2023, ltdd, fairness-acc-tradeoff1}. In this work, we follow existing studies \cite{ltdd, Pessach2023} and use the average of all measure values for each record in our dataset to represent the overall performance. 

\subsubsection{Fairness Measures}
\label{sec:fm}
To evaluate the fairness of recommendation systems, we use two measures that rely on the top-K results and one measure that is independent of the top-K results~\cite{detgreedy, fairness-recsys-survey}. Both measures are useful for analyzing group fairness, which is the goal of our investigation. 
Similar to existing works \cite{corms, revfinder} that conducted their experiments in a specific setting, we limit the values of K to 4, 6, and 10. 


\begin{itemize}
    \item $Skew_\mathrm{S_i}@K$: The skew of the ranked list of top-K candidates for a certain value $S_i$ of the sensitive attribute is:
    
    \begin{equation}
        \label{eq:skewk}
        Skew_\mathrm{S_i}@K(C) = \ln(\frac{P_\mathrm{C^K,S_i}}{P_\mathrm{D,S_i}})
    \end{equation}
    
    where $C$ is the ranked list of candidates, $C^K$ is the top-K candidates from $C$, $P_\mathrm{C^K,S_i}$ is the proportion of candidates having the sensitive attribute value $S_i$ in the top-K results, and $P_\mathrm{D,S_i}$ is the desired proportion of candidates with the sensitive attribute value $S_i$ in the given dataset. The desired proportion for each project is calculated by dividing the number of female reviewers by the total number of reviewers.
    When the proportion of the protected group is smaller than the desired proportion, the fraction's output is less than one, causing the natural logarithm's output to be negative.
    As a consequence, if this measure returns a negative value, we have a fairness problem for the sensitive attribute value $S_i$ and the top-K recommendation list is not a fair recommendation. The ideal value of this measure is zero (since the output of the fraction will be one and the natural logarithm of one is zero). The closer the value of this measure comes to zero, the more fair the recommendation. Yet, if the measure produces a positive result, it indicates that the recommendations favor the protected group \cite{detgreedy}.
    \\

    \item Statistical Parity Difference For top-K results ($SPD@K$): The statistical parity difference (SPD) is a well-known measure of fairness that is used in many articles about how fair machine learning is when it comes to classification tasks \cite{Pessach2023, ltdd}. This measure is demonstrated in Eq. \ref{eq:spd} as an example for the binary classification:

    \begin{equation}
        \label{eq:spd}
        \left| P[\hat{Y} = 1|S = 1] - P[\hat{Y} = 1|S \neq 1]\right| \leq \epsilon
    \end{equation}

    This measure implies that the rate or probability of a positive prediction for the protected group should be similar to that of the privileged group. In the best-case scenario, these probabilities or rates are equal to one another, and the SPD value is zero. Yet in reality, to consider the result of a classification algorithm fair, the difference between these probabilities should be less than a threshold known as epsilon $(\epsilon)$, which should be chosen based on the problem information.

    Nevertheless, in order to use this measure for top-K results in recommendation systems such as code reviewer recommendation systems, we must change the calculation in Eq. \ref{eq:spd}. As a result, we introduce the $SPD@K$ measure, which is described in Eq. \ref{eq:spdk}:

    \begin{equation}
        \label{eq:spdk}
        \left| P[\hat{Y} \in C^K|S = 1] - P[\hat{Y} \in C^K|S \neq 1]\right| \leq \epsilon
    \end{equation}
    
    Where $C^K$ is the ranked list of top-K candidates from the ranked list of candidates $C$ as the result of the recommendation. In this measure, we also refer to $\epsilon$ as the expected SPD threshold. This threshold will be computed by calculating the absolute difference between the male and female ratios in the dataset, which is described in Eq. \ref{eq:expected-spd-threshold}:

     \begin{equation}
        \label{eq:expected-spd-threshold}
        \epsilon =  \left|\frac{\text{\# Females} }{\text{\# Reviewers}} - \frac{\text{\# Males} }{\text{\# Reviewers}}\right|
    \end{equation}
    
    We expect that the difference in recommendation rates for males and females will be similar to the disparity in ratios of these groups in the dataset. \\
    \item Normalized Discounted Cumulative KL-divergence (NDKL)\label{sec:fm-ndkl}: Eq. \ref{eq:ndkl} describes the normalized discounted cumulative Kullback-Leiber (KL) divergence given a ranked list of the candidates $C$:
    \begin{equation}
        \label{eq:ndkl}
        NDKL(C) = \frac{1}{Z} \sum_{i \in Ks} \frac{1}{\log_2 (i + 1)} d_\mathrm{KL}(D_\mathrm{C^i} || D_d)
    \end{equation}
    \\
    In Eq. \ref{eq:ndkl} $d_\mathrm{KL}(D_\mathrm{C^i} || D_d) = \sum_j D_\mathrm{C^i}(j) \ln \frac{D_\mathrm{C^i}(j)}{D_d}$, $Z = \sum_\mathrm{i = 1}^\mathrm{C} \frac{1}{\log_2 (i + 1)}$, and $Ks = \{4, 6, 10\}$ where $D_\mathrm{C^i}$ and $D_d$ represent the proportion of the top $i$ candidates in the ranked list of candidates $C$ having the sensitive attribute $j$, respectively, and the desired proportion based on the dataset with the sensitive attribute value $j$. 

    This measure considers all sensitive attribute values, which means it considers both privileged and underprivileged groups~\cite{detgreedy, fairness-recsys-survey}. It is also independent of the value of K, unlike the other measures, because its value accumulates over different K values. This measure has non-negative values, with zero being the optimal value for a fair recommendation based on it. As a result, if the value is positive, it indicates that the recommendation has fairness issues. The problem with this measure is that we cannot be sure which attribute value is causing the recommendation to be unfair, and it's difficult to interpret. The most effective way to use it is to look for a decrease in the values of this measure, which indicates that fairness has improved~\cite{detgreedy}.
    
\end{itemize}

\subsubsection{Performance Measures}
\label{sec:pa}

In this study, we adopt the identical performance metrics  utilized in previous research papers discussing CORMS and RevFinder \cite{corms, revfinder}, i.e., Top-K Accuracy and  Mean Reciprocal Rank (MRR). This enables us to cross-compare and guarantee accurate replications of both works and to ensure a fair assessment before and after applying bias mitigation strategies. 
Top-K accuracy is a performance measure that shows how often a code reviewer recommendation system can suggest the right code reviewers for a set of reviews \cite{corms, revfinder}. It is calculated by dividing the number of reviews where at least one correct code reviewer was in the top-K recommendations by the total number of reviews. 
MRR is defined as the average of the reciprocal ranks of the correct answers in the ranked list of recommended code reviewers \cite{corms, revfinder}. In other words, for each review, MRR calculates the reciprocal rank of the first relevant result and averages these values over all the reviews in the dataset. 
    %
In our scenario, we use $MRR@K$ to focus on the top-K recommendations. This means that we only consider the top-K candidates in the original calculations. Every candidate not on the top-K list has a reciprocal rank of zero. The rest of the calculations remain the same.
    

\subsection{Selected Unfairness Mitigation Approaches}
\label{sec:sfa}


Researchers have proposed several unfairness mitigation strategies for recommendation systems~\cite{fairness-recsys-survey, fa-ir, user-oriented-fairness, detgreedy, amortizing-individual-fairness, cpfair, fairrec}, but not all of them are applicable to our case due to different views on fairness. Also, due to the inherent differences in the fairness of classification and recommendation tasks (e.g., techniques used for classification do not support top-K results), 
we cannot employ those mitigation techniques for classification tasks in our study. 

In this work we select two applicable approaches from the ``post-processing'' category of fairness mitigation methods (see \ref{sec:ma}) for reviewer recommendation systems~\cite{fairness-recsys-survey}. 
We did not apply any ``in-processing'' mitigation approaches as these approaches require non-trivial changes to the code reviewer recommendation systems, which do not generalize. In addition, we exclude all the ``pre-processing'' mitigation techniques, as these data-oriented techniques in the literature are mostly designed for classification tasks, thus are not suitable for our study~\cite{fairness-recsys-preproc1, fairness-recsys-preproc2}. 
The details of these two mitigation approaches are as follows. 

\subsubsection{DetGreedy Algorithm}
Geyik et al. \cite{detgreedy} developed this algorithm for fairness-aware recommendation in LinkedIn Talent Search recommendation systems. This algorithm works as follows.
Given a top-K list, there are two requirements to satisfy the fairness condition:
\begin{enumerate}
    \item[a.] Min: $\forall K < \left| C \right| \& \forall s_i \in A, count_K(S_i) \ge \left\lfloor P_{D,S_i} \cdot K \right\rfloor$
    \item[b.] Max: $\forall K < \left| C \right| \& \forall s_i \in A, count_K(S_i) \le \left\lceil P_{D,S_i} \cdot K \right\rceil$
\end{enumerate}
Where $A$ is the set of attribute values, $C$ is the list of candidates, $P_{D,S_i}$ is the desired proportion of candidates with the attribute value $S_i$, and $count_k(S_i)$ is the number of candidates with the attribute value $S_i$ in the top-K results.
If there are candidates whose attribute values are close to violating the minimum requirement, choose the candidate with the highest score next from that group. If all candidates meet the minimum requirement, then pick the one with the highest next score from the group based on the attribute value among those who have not yet reached their maximum requirements. 


\subsubsection{DetRelaxed}
To improve DetGreedy, Geyik et al.~\cite{detgreedy} further proposed DetRelaxed. 
While DetGreedy aims to include as many high-scoring candidates as possible in the ranked list, it may not be effective in various scenarios, according to the authors \cite{detgreedy}. The DetRelaxed algorithm was proposed to consider all candidates who satisfy the minimum requirement and minimize the term $\left\lceil \frac{\left\lceil P_{D,S_i} \cdot K \right\rceil}{P_{D,S_i}} \right\rceil$ to select the candidate with the highest score for the next position.

According to the previous study \cite{detgreedy}, these approaches resulted in a huge improvement in fairness at the production stage. Hence, we select these approaches to assess their effectiveness with our subjects.






\label{sec:pa}


\section{Results and Analysis}
\label{sec:result}
This section presents the experimental results and answers the research questions we asked in Section~\ref{sec:rqs}.



\subsection{RQ1: Existence of Fairness}
\label{sec:answer_rq1}

\noindent \textbf{Approach:} 
To answer this RQ, we first build RevFinder and CORMS on the training data selected from our experimental dataset (details are in Section~\ref{sec:dc}). 
When training each model, following existing studies~\cite{revfinder,corms}, we divide the dataset into two sections, with 80\% for training and the remaining 20\% for testing, chronologically. 
Then, we use the measures mentioned in Section~\ref{sec:fm} to evaluate the fairness of the recommendations generated by these code reviewer recommendation systems based on the testing data. 

\noindent \textbf{Result:}  
Table~\ref{table:final-results} presents the findings of our experiments, where we analyze all the evaluation measures for each project across three different scenarios, i.e., original (i.e., results obtained without any mitigation approaches), DG (i.e., outcomes obtained after using DetGreedy), and DR (i.e., results obtained after using DetRelaxed). Also, it presents the SPD threshold for each project in the dataset. 
As we can see from column ``Original'' in the table, $Skew@K$ has negative values under both CORMS and RevFinder on the four projects with different K values, e.g., BSSW, GetSentry, Node.js, and Shopify, which have negative $Skew@K$ values, i.e, $-0.19$, $-0.01$, $-0.73$, and $-0.90$ under CORMS when recommending top 10 reviewers. 
For $SPD@K$, we show the detailed desired SPD threshold of each project in the last row in Table~\ref{table:final-results}. As we can see on most projects, the values of $SPD@K$ are above the specified threshold values. For the projects BSSW, GetSentry, Node.js, and Shopify, the variation between the $SPD@K$ measure values and the SPD threshold is $0.15$, $0.03$, $0.03$, and $0.08$, respectively. This indicates that the disparity between the percentages of males and females being recommended in BSSW, GetSentry, Node.js, and Shopify projects is $15\%$, $3\%$, $3\%$, and $8\%$, respectively. 
As an illustration, in the Shopify project's Top-6 scenario, the $SPD@K$ measures for both CORMS and RevFinder are 0.80 and 0.97 and are highlighted in yellow, respectively, which exceeded the anticipated SPD threshold of 0.79, indicating that the results were unfair.
Male reviewers get an average of 7.25\% more recommendations than female code reviewers compared to their distribution in the reviewer set. 
Results from $Skew@K$ and $SPD@K$ indicate there are unfairness issues for each of the four experimental projects under both CORMS and RevFinder. 


Furthermore, we analyze the normalized discounted KL divergence measure (NDKL), which is shown in Table~\ref{table:final-results}. Our experiments indicate that the NDKL measure yielded positive values for all projects, suggesting that the outcomes may be biased towards a particular value of the sensitive attribute. As discussed in Section~\ref{sec:fm-ndkl}, a disadvantage of this measure is its challenging interpretation due to the unknown sensitive attribute value that causes unfairness. However, we acknowledge that the recommendation could be unfair, which helps us address this research question.

    
\mybox{\textbf{Answer to RQ1:} In the exmained projects, male reviewers get an average of 7.25\% more recommendations than female code reviewers, considering their distribution in the recommendation list under both CORMS and RevFinder. Our experiment results suggest that the two ML-based code reviewer recommendation systems studied exhibit bias towards gender.} 

\subsection{RQ2: Root Cause for Unfairness}
\label{sec:answer_rq2}

\begin{table*}
\centering
\caption{Summary of experiment results. The results that are considered unfair recommendations based on the $SPD@K$ and $Skew@K$ measures are highlighted in \tikzcircle[yellow, fill=yellow]{2pt} and \tikzcircle[orange, fill=orange]{2pt}, respectively.
}
\label{table:final-results}
\resizebox{\textwidth}{!}{
\begin{tabular}{|c|c|c|ccc|ccc|ccc|ccc|} 
\hline
\multirow{3}{*}{\textbf{Top-K}} & \multirow{3}{*}{\textbf{Subjects}} & \multirow{3}{*}{\textbf{Measures}} & \multicolumn{12}{c|}{\textbf{Projects}}                                                                                                                                                                                                                                                                                                                                                                                                                                                                                         \\ 
\cline{4-15}
                                &                                    &                                    & \multicolumn{3}{c|}{\textbf{BSSW}}                                                           & \multicolumn{3}{c|}{\textbf{GetSentry}}                                                                                            & \multicolumn{3}{c|}{\textbf{Node.js}}                                                                                                            & \multicolumn{3}{c|}{\textbf{Shopify}}                                                                                                    \\ 
\cline{4-15}
                                &                                    &                                    & \multicolumn{1}{c|}{\textbf{Original}}    & \multicolumn{1}{c|}{\textbf{DG}} & \textbf{DR}   & \multicolumn{1}{c|}{\textbf{Original}}   & \multicolumn{1}{c|}{\textbf{DG}}             & \textbf{DR}                              & \multicolumn{1}{c|}{\textbf{Original}}   & \multicolumn{1}{c|}{\textbf{DG}}                  & \textbf{DR}                                       & \multicolumn{1}{c|}{\textbf{Original}}   & \multicolumn{1}{c|}{\textbf{DG}}                  & \textbf{DR}                               \\ 
\hline
\multirow{8}{*}{Top-4}          & \multirow{4}{*}{CORMS}             & TopK-ACC                           & 79\%                                      & 79\%                             & 79\%          & 43\%                                     & 43\%                                         & 37\%                                     & 41\%                                     & 41\%                                              & 41\%                                              & 25\%                                     & 25\%                                              & 30\%                                      \\
                                &                                    & MRR@K                              & 0.54                                      & 0.54                             & 0.54          & 0.23                                     & 0.23                                         & 0.21                                     & 0.26                                     & 0.26                                              & 0.27                                              & 0.19                                     & 0.19                                              & 0.21                                      \\
                                &                                    & SPD@K                              & 0                                         & 0                                & 0             & {\cellcolor[rgb]{1,0.988,0.62}}0.93      & {\cellcolor[rgb]{1,0.988,0.62}}0.93          & {\cellcolor[rgb]{1,0.988,0.62}}1         & {\cellcolor[rgb]{1,0.988,0.62}}0.85      & {\cellcolor[rgb]{1,0.988,0.62}}0.85               & {\cellcolor[rgb]{1,0.988,0.62}}1                  & {\cellcolor[rgb]{1,0.988,0.62}}0.80      & {\cellcolor[rgb]{1,0.988,0.62}}0.87               & {\cellcolor[rgb]{1,0.988,0.62}}1          \\
                                &                                    & Skew@K                             & 0.01                                      & 0.01                             & 0.01          & {\cellcolor[rgb]{0.988,0.82,0.514}}-1.82 & {\cellcolor[rgb]{0.988,0.82,0.514}}-1.82     & {\cellcolor[rgb]{0.988,0.82,0.514}}-2.25 & {\cellcolor[rgb]{0.988,0.82,0.514}}-1.79 & {\cellcolor[rgb]{0.988,0.82,0.514}}-1.79          & {\cellcolor[rgb]{0.988,0.82,0.514}}-2.7           & {\cellcolor[rgb]{0.988,0.82,0.514}}-1.41 & {\cellcolor[rgb]{0.988,0.82,0.514}}-1.50          & {\cellcolor[rgb]{0.988,0.82,0.514}}-2.35  \\ 
\cline{2-15}
                                & \multirow{4}{*}{RevFinder}         & TopK-ACC                           & 71\%                                      & 73\%                             & 73\%          & 43\%                                     & 43\%                                         & 37\%                                     & 50\%                                     & 50\%                                              & 45\%                                              & 25\%                                     & 25\%                                              & 24\%                                      \\
                                &                                    & MRR@K                              & 0.5                                       & 0.5                              & 0.5           & 0.25                                     & 0.25                                         & 0.22                                     & 0.32                                     & 0.32                                              & 0.31                                              & 0.15                                     & 0.15                                              & 0.14                                      \\
                                &                                    & SPD@K                              & {\cellcolor[rgb]{1,0.988,0.62}}0.09       & \textbf{0}                       & \textbf{0}    & {\cellcolor[rgb]{1,0.988,0.62}}0.90      & {\cellcolor[rgb]{1,0.988,0.62}}\textbf{0.88} & {\cellcolor[rgb]{1,0.988,0.62}}1         & 0.52                                     & 0.56                                              & {\cellcolor[rgb]{1,0.988,0.62}}1                  & {\cellcolor[rgb]{1,0.988,0.62}}0.97      & {\cellcolor[rgb]{1,0.988,0.62}}\textbf{0.96}      & {\cellcolor[rgb]{1,0.988,0.62}}1          \\
                                &                                    & Skew@K                             & {\cellcolor[rgb]{0.988,0.82,0.514}}-0.001 & \textbf{0.01}                    & \textbf{0.01} & {\cellcolor[rgb]{0.988,0.82,0.514}}-1.48 & {\cellcolor[rgb]{0.988,0.82,0.514}}-1.48     & -2.12                                    & 0.15                                     & \textbf{0.09}                                     & {\cellcolor[rgb]{0.988,0.82,0.514}}-2.71          & {\cellcolor[rgb]{0.988,0.82,0.514}}-2.03 & {\cellcolor[rgb]{0.988,0.82,0.514}}\textbf{-1.94} & {\cellcolor[rgb]{0.988,0.82,0.514}}-2.17  \\ 
\hline
\multirow{8}{*}{Top-6}          & \multirow{4}{*}{CORMS}             & TopK-ACC                           & 84\%                                      & 87\%                             & 87\%          & 44\%                                     & 44\%                                         & 38\%                                     & 52\%                                     & 52\%                                              & 47\%                                              & 34\%                                     & 36\%                                              & 37\%                                      \\
                                &                                    & MRR@K                              & 0.55                                      & 0.56                             & 0.56          & 0.24                                     & 0.24                                         & 0.21                                     & 0.29                                     & 0.29                                              & 0.28                                              & 0.21                                     & 0.21                                              & 0.22                                      \\
                                &                                    & SPD@K                              & {\cellcolor[rgb]{1,0.988,0.62}}0.31       & \textbf{0}                       & \textbf{0}    & {\cellcolor[rgb]{1,0.988,0.62}}0.87      & {\cellcolor[rgb]{1,0.988,0.62}}0.87          & {\cellcolor[rgb]{1,0.988,0.62}}1         & {\cellcolor[rgb]{1,0.988,0.62}}0.86      & {\cellcolor[rgb]{1,0.988,0.62}}0.86               & \textbf{0.66}                                     & {\cellcolor[rgb]{1,0.988,0.62}}0.80      & {\cellcolor[rgb]{1,0.988,0.62}}0.87               & {\cellcolor[rgb]{1,0.988,0.62}}1          \\
                                &                                    & Skew@K                             & {\cellcolor[rgb]{0.988,0.82,0.514}}-0.35  & \textbf{0.01}                    & \textbf{0.01} & {\cellcolor[rgb]{0.988,0.82,0.514}}-1.20 & {\cellcolor[rgb]{0.988,0.82,0.514}}-1.20     & {\cellcolor[rgb]{0.988,0.82,0.514}}-2.25 & {\cellcolor[rgb]{0.988,0.82,0.514}}-1.56 & {\cellcolor[rgb]{0.988,0.82,0.514}}-1.56          & \textbf{0.12}                                     & {\cellcolor[rgb]{0.988,0.82,0.514}}-1.14 & {\cellcolor[rgb]{0.988,0.82,0.514}}-1.28          & {\cellcolor[rgb]{0.988,0.82,0.514}}-2.35  \\ 
\cline{2-15}
                                & \multirow{4}{*}{RevFinder}         & TopK-ACC                           & 88\%                                      & 83\%                             & 83\%          & 49\%                                     & 49\%                                         & 46\%                                     & 54\%                                     & 54\%                                              & 54\%                                              & 36\%                                     & 36\%                                              & 37\%                                      \\
                                &                                    & MRR@K                              & 0.53                                      & 0.52                             & 0.52          & 0.26                                     & 0.26                                         & 0.24                                     & 0.33                                     & 0.33                                              & 0.33                                              & 0.17                                     & 0.17                                              & 0.17                                      \\
                                &                                    & SPD@K                              & {\cellcolor[rgb]{1,0.988,0.62}}0.13       & \textbf{0}                       & \textbf{0}    & 0.81                                     & \textbf{0.80}                                & {\cellcolor[rgb]{1,0.988,0.62}}1         & 0.60                                     & 0.69                                              & 0.66                                              & {\cellcolor[rgb]{1,0.988,0.62}}0.97      & {\cellcolor[rgb]{1,0.988,0.62}}\textbf{0.96}      & {\cellcolor[rgb]{1,0.988,0.62}}1          \\
                                &                                    & Skew@K                             & {\cellcolor[rgb]{0.988,0.82,0.514}}-0.14  & \textbf{0.01}                    & \textbf{0.01} & {\cellcolor[rgb]{0.988,0.82,0.514}}-0.53 & {\cellcolor[rgb]{0.988,0.82,0.514}}-0.58     & {\cellcolor[rgb]{0.988,0.82,0.514}}-2.12 & 0.07                                     & {\cellcolor[rgb]{0.988,0.82,0.514}}-0.1           & 0.15                                              & {\cellcolor[rgb]{0.988,0.82,0.514}}-1.97 & {\cellcolor[rgb]{0.988,0.82,0.514}}\textbf{-1.87} & {\cellcolor[rgb]{0.988,0.82,0.514}}-2.17  \\ 
\hline
\multirow{8}{*}{Top-10}         & \multirow{4}{*}{CORMS}             & TopK-ACC                           & 94\%                                      & 100\%                            & 100\%         & 45\%                                     & 46\%                                         & 45\%                                     & 58\%                                     & 58\%                                              & 58\%                                              & 42\%                                     & 43\%                                              & 46\%                                      \\
                                &                                    & MRR@K                              & 0.57                                      & 0.57                             & 0.57          & 0.24                                     & 0.24                                         & 0.24                                     & 0.29                                     & 0.29                                              & 0.29                                              & 0.22                                     & 0.22                                              & 0.22                                      \\
                                &                                    & SPD@K                              & {\cellcolor[rgb]{1,0.988,0.62}}0.2        & \textbf{0}                       & \textbf{0}    & 0.79                                     & 0.82                                         & 0.8                                      & {\cellcolor[rgb]{1,0.988,0.62}}0.82      & {\cellcolor[rgb]{1,0.988,0.62}}\textbf{0.78}      & {\cellcolor[rgb]{1,0.988,0.62}}\textbf{0.8}       & 0.77                                     & \textbf{0.73}                                     & {\cellcolor[rgb]{1,0.988,0.62}}0.8        \\
                                &                                    & Skew@K                             & {\cellcolor[rgb]{0.988,0.82,0.514}}-0.19  & \textbf{0.01}                    & \textbf{0.01} & {\cellcolor[rgb]{0.988,0.82,0.514}}-0.01 & {\cellcolor[rgb]{0.988,0.82,0.514}}-0.1      & \textbf{0.14}                            & {\cellcolor[rgb]{0.988,0.82,0.514}}-0.73 & {\cellcolor[rgb]{0.988,0.82,0.514}}\textbf{-0.31} & {\cellcolor[rgb]{0.988,0.82,0.514}}\textbf{-0.35} & {\cellcolor[rgb]{0.988,0.82,0.514}}-0.90 & \textbf{0.26}                                     & \textbf{0.04}                             \\ 
\cline{2-15}
                                & \multirow{4}{*}{RevFinder}         & TopK-ACC                           & 88\%                                      & 92\%                             & 92\%          & 62\%                                     & 62\%                                         & 55\%                                     & 68\%                                     & 68\%                                              & 68\%                                              & 44\%                                     & 44\%                                              & 43\%                                      \\
                                &                                    & MRR@K                              & 0.53                                      & 0.53                             & 0.53          & 0.28                                     & 0.28                                         & 0.25                                     & 0.35                                     & 0.35                                              & 0.35                                              & 0.18                                     & 0.18                                              & 0.17                                      \\
                                &                                    & SPD@K                              & {\cellcolor[rgb]{1,0.988,0.62}}0.17       & \textbf{0}                       & \textbf{0}    & 0.81                                     & 0.82                                         & {\cellcolor[rgb]{1,0.988,0.62}}1         & 0.68                                     & 0.68                                              & {\cellcolor[rgb]{1,0.988,0.62}}0.8                & {\cellcolor[rgb]{1,0.988,0.62}}0.96      & {\cellcolor[rgb]{1,0.988,0.62}}0.96               & {\cellcolor[rgb]{1,0.988,0.62}}1          \\
                                &                                    & Skew@K                             & {\cellcolor[rgb]{0.988,0.82,0.514}}-0.19  & \textbf{0.01}                    & \textbf{0.01} & {\cellcolor[rgb]{0.988,0.82,0.514}}-0.03 & {\cellcolor[rgb]{0.988,0.82,0.514}}-0.16     & {\cellcolor[rgb]{0.988,0.82,0.514}}-2.12 & 0.06                                     & 0.06                                              & {\cellcolor[rgb]{0.988,0.82,0.514}}-0.32          & {\cellcolor[rgb]{0.988,0.82,0.514}}-1.83 & {\cellcolor[rgb]{0.988,0.82,0.514}}\textbf{-1.77} & {\cellcolor[rgb]{0.988,0.82,0.514}}-2.17  \\ 
\hline
\multicolumn{1}{c|}{}           & CORMS                              & NDKL                               & 0.04                                      & \textbf{0.01}                    & \textbf{0.01} & 0.08                                     & 0.08                                         & 0.08                                     & 0.11                                     & \textbf{0.10}                                     & \textbf{0.08}                                     & 0.14                                     & \textbf{0.08}                                     & \textbf{0.09}                             \\ 
\cline{2-2}
\multicolumn{1}{c|}{}           & RevFinder                          & NDKL                               & 0.04                                      & \textbf{0.01}                    & \textbf{0.01} & 0.07                                     & 0.07                                         & 0.09                                     & 0.06                                     & 0.06                                              & 0.08                                              & 0.09                                     & 0.09                                              & 0.10                                      \\ 
\cline{2-15}
\multicolumn{1}{c|}{}           & \multicolumn{2}{c|}{SPD Threshold}                                      & \multicolumn{3}{c|}{0}                                                                       & \multicolumn{3}{c|}{0.82}                                                                                                          & \multicolumn{3}{c|}{0.69}                                                                                                                        & \multicolumn{3}{c|}{0.79}                                                                                                                \\
\cline{2-15}
\end{tabular}
}
\label{table:final-results}
\end{table*}

\noindent \textbf{Approach:} 
To answer this question, we first conduct an exclusive analysis of current fairness studies in the literature to collect all the possible factors that have been examined to be effective in determining the fairness of ML applications~\cite{bias-fairness-survey2021, bias-recsys-survey, Pessach2023, popularity-bias, fairness-recsys-survey}, especially recommendation systems. Four factors were collected: (1) imbalanced or skewed data, (2) popularity bias, and (3) algorithmic objectives. 
We are not going to examine algorithmic objectives since it requires us to examine the code reviewer recommendation architectures, which is out of the scope of this paper. As a result, we employ the factors of imbalanced or skewed data and popularity bias to explore the possible root causes of code reviewer recommendation systems.

Imbalanced or skewed data can lead to unfairness because models trained on such data have a strong probability of learning behavior towards over-represented groups, eventually becoming overfitted to them \cite{fairness-recsys-survey, bias-recsys-survey}. Popularity bias and unfairness also have a strong connection with each other \cite{bias-recsys-survey, popularity-bias}. The ``long-tail effect'' occurs when recommendation systems favor popular items over less popular ones \cite{longtail1, longtail2}, which leads to discrimination against the less popular items. If the less popular items are generally from the protected group, popularity bias will turn into unfairness.




\noindent \textbf{Result:} The following are the primary factors responsible for the unfairness and disparities between actual and expected distributions of the protected group in the outputs of the code reviewer recommendation system:

\textbf{Imbalanced or Skewed Data}: Our analysis indicates that the imbalanced representation of male and female reviewers in some projects is one of the reasons behind the unfair outcomes of code reviewer recommendation systems. Table \ref{table:datasets} shows that, apart from the BSSW project, which has an equal number of female and male reviewers, the number of male reviewers is higher than that of female reviewers in all other projects (males are approximately 8 times more than females), which leads to skewed data. The BSSW project has a record of females accounting for 67\% of the total, while in the Node.js, GetSentry, and Shopify projects, this percentage is 13\%, 8\%, and 6\%, respectively. 
Consequently, Table \ref{table:final-results} reveals that the results of the $SPD@K$ measure for the BSSW project are substantially lower than those of other projects, suggesting a fairer outcome. The same pattern is observed for the $Skew@K$ measure in the BSSW project, as other projects exhibit more negative values for this measure, indicating larger discrepancies between the current and fair conditions.
    

\textbf{Popularity Bias}: In the projects examined, bias was found to be more prevalent in projects where male reviewers were more popular than female reviewers. For instance, in Shopify and Node.js, the first 14\% and 15\% of popular reviewers were all male, respectively. This led to unfairness in the CORMS and RevFinder recommendation systems, which learned and incorporated this preference for male reviewers into their decision-making processes. In contrast, in the BSSW project, the first two popular reviewers were female, resulting in fairness measures like $Skew@K$ and $SPD@K$ being closer to fair values, such as $Skew@K$ being closer to zero. More details can be found in the supplement materials.

\mybox{\textbf{Answer to RQ2:} We confirm that popularity bias and imbalanced or skewed data are two factors that can affect code reviewer recommendation systems, both of which can lead to unfairness. Projects that do not have these issues (e.g., BSSW) result in values in fairness measures that are closer to a fair state, in contrast to projects that do suffer from these problems.}

\subsection{RQ3: Effectiveness of Existing Unfairness Mitigation Techniques}
\label{sec:answer_rq3}

\noindent \textbf{Approach:} In this RQ, we examine the selected unfairness mitigation approaches (i.e., DetGreedy and DetRelaxed) in Section \ref{sec:sfa} to see if they could improve the fairness of code reviewer recommendation systems. Furthermore, because applying an unfairness mitigation technique has been shown to have a trade-off with performance measures \cite{fairness-acc-tradeoff1, fairness-acc-tradeoff2}, we should also guarantee that applying these mechanisms does not adversely affect performance measures. As a result, we compare performance measures before and after using unfairness mitigation techniques. 

\noindent 
 \textbf{Result:} 
The fairness measures after applying these two mitigation approaches, i.e., DetGreedy and DetRelaxed, for each project, are shown in Table~\ref{table:final-results}.

As we can see from the table, both DetGreedy and DetRelaxed mitigation techniques can improve fairness, but not across all projects. The bolded values on this table indicate fairness improvements compared to original settings. 
With the use of these approaches, the BSSW project saw a significant fairness improvement of 100\%. Also, in the BSSW project, RevFinder's top-K accuracy decreased by 5\% solely in the top-6 scenario. 
While performance measures were not adversely impacted (only a 1.75\% decrease on average on all projects), the fairness enhancement was not as noticeable in projects with an imbalanced distribution of male and female reviewers (e.g., Node.js, GetSentry, and Shopify). 

\mybox{\textbf{Answer to RQ3:} DetGreedy and DetRelaxed mitigation approaches can improve fairness while maintaining performance, but not consistently across all projects.}



\section{Is it possible to further improve the fairness issue in code reviewer recommendation systems?}
\label{sec:discussion}




As presented in Section \ref{sec:answer_rq3}, although both examined mitigation approaches, DetGreedy and DetRelaxed, can help improve fairness for some projects, we can also see that both DetGreedy and DetRelaxed~\cite{detgreedy} do not work effectively for certain projects. For example, the fairness improvement in the BSSW project is significantly higher than that of other projects e.g., Node.js, GetSentry, and Shopify. 
Our analysis shows that these two approaches can be effective when the data is not imbalanced or skewed in terms of protected and privileged group ratios. However, for projects with imbalanced or skewed data, these approaches may not be effective in improving fairness. As highlighted in Section \ref{sec:answer_rq2}, imbalanced or skewed data is one of the main factors causing unfairness in such systems. Therefore, a solution that can address this issue is necessary for improving fairness. 

\begin{algorithm}
\caption{Improved Greedy Re-Ranking Algorithm} 
\begin{algorithmic}[1]
\Procedure{IGRR}{scores\_list, SPD\_threshold, K}

\State topK\_list = topK(scores\_list, K)
\State topK\_excluded = exclude\_topK(scores\_list)
\State other\_females = get\_females(topK\_excluded)
\State other\_males = get\_males(topK\_excluded)

\While{is\_unfair(SPD(topK\_list), SPD\_threshold, K)}
    \State before\_scores\_list = topK\_list
    \State m\_val\_before = metric(topK\_list)
    \If{unfair\_towards\_males(topK\_list)}
        \State substitute(topK\_list, other\_females)
    \Else
        \State substitute(topK\_list, other\_males)
    \EndIf
    \State m\_val\_after = metric(topK\_list)
    \If{m\_val\_after $\geq$ m\_val\_before}
        \State \Return{sort(before\_scores\_list)}
    \EndIf
\EndWhile
\State \Return{sort(topK\_list)}

\EndProcedure
\end{algorithmic}
\label{algorithm1}
\end{algorithm}

In order to tackle this problem, we propose an improved greedy re-ranking algorithm. The motivation behind this algorithm is that it should guarantee to at least bring a reviewer from the discriminated group into the top-K list, regardless of other constraints. Although the DetGreedy and DetRelaxed algorithms specify the minimum and maximum requirements in Section \ref{sec:sfa}, following both requirements as a fairness condition does not necessarily result in this outcome. 
Algorithm \ref{algorithm1} demonstrates the algorithm of our approach. The method aims to improve fairness in a recommendation list by using the $SPD@K$ measure. It replaces a female reviewer with the highest score not on the top-K list with a male reviewer with the lowest score on the top-K list to achieve the SPD threshold. This process is repeated until the threshold is met.

\begin{table}[]
\caption{Results of fairness mitigation effectiveness of our approach. \tikzcircle[green, fill=green]{2pt} indicates the results where our approach works but the DetGreedy and DetRelaxed methods did not succeed in achieving fairness improvements. \tikzcircle[cyan, fill=cyan]{2pt} indicates that where our approach outperformed the DetGreedy and DetRelaxed approaches in terms of fairness improvements. The unfair recommendations based on the $SPD@K$ and $Skew@K$ measures are highlighted in \tikzcircle[yellow, fill=yellow]{2pt} and \tikzcircle[orange, fill=orange]{2pt}, respectively.}
\resizebox{9cm}{!}{
\begin{tabular}{c|cc|c|c|c|c|}
\hline
\multicolumn{1}{|c|}{}                                 & \multicolumn{1}{c|}{}                                    &                                     &                                 &                                       &                                       &                                       \\
\multicolumn{1}{|c|}{\multirow{-2}{*}{\textbf{Top-K}}} & \multicolumn{1}{c|}{\multirow{-2}{*}{\textbf{Subjects}}} & \multirow{-2}{*}{\textbf{Measures}} & \multirow{-2}{*}{\textbf{BSSW}} & \multirow{-2}{*}{\textbf{GetSentry}}  & \multirow{-2}{*}{\textbf{Node.js}}    & \multirow{-2}{*}{\textbf{Shopify}}    \\ \hline
\multicolumn{1}{|c|}{}                                 & \multicolumn{1}{c|}{}                                    & Top-K-ACC                           & 53\%                            & 40\%                                  & 41\%                                  & 24\%                                  \\
\multicolumn{1}{|c|}{}                                 & \multicolumn{1}{c|}{}                                    & MRR@K                               & 0.43                            & 0.23                                  & 0.26                                  & 0.19                                  \\
\multicolumn{1}{|c|}{}                                 & \multicolumn{1}{c|}{}                                    & SPD@K                               & \cellcolor[HTML]{FFFE65}0.5     & \cellcolor[HTML]{9AFF99}\textbf{0.5}  & \cellcolor[HTML]{9AFF99}\textbf{0.5}  & \cellcolor[HTML]{9AFF99}\textbf{0.43} \\
\multicolumn{1}{|c|}{}                                 & \multicolumn{1}{c|}{\multirow{-4}{*}{CORMS}}             & Skew@K                              & \cellcolor[HTML]{FFCE93}-0.65   & \cellcolor[HTML]{9AFF99}\textbf{1}    & \cellcolor[HTML]{9AFF99}\textbf{0.5}  & \cellcolor[HTML]{9AFF99}\textbf{1}    \\ \cline{2-7} 
\multicolumn{1}{|c|}{}                                 & \multicolumn{1}{c|}{}                                    & Top-K-ACC                           & 57\%                            & 39\%                                  & 50\%                                  & 21\%                                  \\
\multicolumn{1}{|c|}{}                                 & \multicolumn{1}{c|}{}                                    & MRR@K                               & 0.42                            & 0.24                                  & 0.32                                  & 0.13                                  \\
\multicolumn{1}{|c|}{}                                 & \multicolumn{1}{c|}{}                                    & SPD@K                               & \cellcolor[HTML]{FFFE65}0.5     & \cellcolor[HTML]{96FFFB}\textbf{0.5}  & \cellcolor[HTML]{9AFF99}\textbf{0.45} & \cellcolor[HTML]{96FFFB}\textbf{0.5}  \\
\multicolumn{1}{|c|}{\multirow{-8}{*}{Top-4}}          & \multicolumn{1}{c|}{\multirow{-4}{*}{RevFinder}}         & Skew@K                              & \cellcolor[HTML]{FFCE93}-0.55   & \cellcolor[HTML]{9AFF99}\textbf{1.13} & \cellcolor[HTML]{96FFFB}\textbf{0.6}  & \cellcolor[HTML]{96FFFB}\textbf{1.07} \\ \hline
\multicolumn{1}{|c|}{}                                 & \multicolumn{1}{c|}{}                                    & Top-K-ACC                           & 87\%                            & 43\%                                  & 47\%                                  & 32\%                                  \\
\multicolumn{1}{|c|}{}                                 & \multicolumn{1}{c|}{}                                    & MRR@K                               & 0.56                            & 0.24                                  & 0.28                                  & 0.2                                   \\
\multicolumn{1}{|c|}{}                                 & \multicolumn{1}{c|}{}                                    & SPD@K                               & \cellcolor[HTML]{FFFE65}0.33    & \cellcolor[HTML]{9AFF99}\textbf{0.66} & \textbf{0.66}                         & \cellcolor[HTML]{9AFF99}\textbf{0.59} \\
\multicolumn{1}{|c|}{}                                 & \multicolumn{1}{c|}{\multirow{-4}{*}{CORMS}}             & Skew@K                              & \textbf{0.26}                   & \cellcolor[HTML]{9AFF99}\textbf{0.61} & \textbf{0.12}                         & \cellcolor[HTML]{9AFF99}\textbf{0.66} \\ \cline{2-7} 
\multicolumn{1}{|c|}{}                                 & \multicolumn{1}{c|}{}                                    & Top-K-ACC                           & 76\%                            & 49\%                                  & 54\%                                  & 33\%                                  \\
\multicolumn{1}{|c|}{}                                 & \multicolumn{1}{c|}{}                                    & MRR@K                               & 0.51                            & 0.26                                  & 0.33                                  & 0.16                                  \\
\multicolumn{1}{|c|}{}                                 & \multicolumn{1}{c|}{}                                    & SPD@K                               & \cellcolor[HTML]{FFFE65}0.33    & \cellcolor[HTML]{96FFFB}\textbf{0.66} & \cellcolor[HTML]{9AFF99}\textbf{0.57} & \cellcolor[HTML]{96FFFB}\textbf{0.66} \\
\multicolumn{1}{|c|}{\multirow{-8}{*}{Top-6}}          & \multicolumn{1}{c|}{\multirow{-4}{*}{RevFinder}}         & Skew@K                              & \cellcolor[HTML]{FFCE93}-0.1    & \cellcolor[HTML]{9AFF99}\textbf{0.75} & \cellcolor[HTML]{9AFF99}\textbf{0.33} & \cellcolor[HTML]{96FFFB}\textbf{0.69} \\ \hline
\multicolumn{1}{|c|}{}                                 & \multicolumn{1}{c|}{}                                    & Top-K-ACC                           & 97\%                            & 45\%                                  & 58\%                                  & 40\%                                  \\
\multicolumn{1}{|c|}{}                                 & \multicolumn{1}{c|}{}                                    & MRR@K                               & 0.57                            & 0.24                                  & 0.29                                  & 0.22                                  \\
\multicolumn{1}{|c|}{}                                 & \multicolumn{1}{c|}{}                                    & SPD@K                               & \cellcolor[HTML]{FFFE65}0.2     & \cellcolor[HTML]{9AFF99}\textbf{0.77} & \cellcolor[HTML]{96FFFB}\textbf{0.6}  & \cellcolor[HTML]{96FFFB}\textbf{0.53} \\
\multicolumn{1}{|c|}{}                                 & \multicolumn{1}{c|}{\multirow{-4}{*}{CORMS}}             & Skew@K                              & \textbf{0.19}                   & \cellcolor[HTML]{9AFF99}\textbf{0.23} & \cellcolor[HTML]{96FFFB}\textbf{0.29} & \cellcolor[HTML]{96FFFB}\textbf{0.81} \\ \cline{2-7} 
\multicolumn{1}{|c|}{}                                 & \multicolumn{1}{c|}{}                                    & Top-K-ACC                           & 88\%                            & 62\%                                  & 59\%                                  & 44\%                                  \\
\multicolumn{1}{|c|}{}                                 & \multicolumn{1}{c|}{}                                    & MRR@K                               & 0.53                            & 0.28                                  & 0.34                                  & 0.18                                  \\
\multicolumn{1}{|c|}{}                                 & \multicolumn{1}{c|}{}                                    & SPD@K                               & \cellcolor[HTML]{FFFE65}0.2     & \cellcolor[HTML]{9AFF99}\textbf{0.78} & \cellcolor[HTML]{9AFF99}\textbf{0.6}  & \cellcolor[HTML]{9AFF99}\textbf{0.79} \\
\multicolumn{1}{|c|}{\multirow{-8}{*}{Top-10}}         & \multicolumn{1}{c|}{\multirow{-4}{*}{RevFinder}}         & Skew@K                              & \textbf{0}                      & \cellcolor[HTML]{9AFF99}\textbf{0.31} & \cellcolor[HTML]{9AFF99}\textbf{0.32} & \cellcolor[HTML]{96FFFB}\textbf{0.22} \\ \hline
                                                       & \multicolumn{1}{c|}{CORMS}                               & NDKL                                & 0.09                            & \textbf{0.06}                         & \textbf{0.03}                         & \textbf{0.11}                         \\ \cline{2-2}
                                                       & \multicolumn{1}{c|}{RevFinder}                           & NDKL                                & 0.09                            & \textbf{0.08}                         & \textbf{0.05}                         & \textbf{0.07}                         \\ \cline{2-7} 
                                                       & \multicolumn{2}{c|}{SPD Threshold}                                                             & 0                               & 0.82                                  & 0.69                                  & 0.79                                  \\ \cline{2-7} 
\end{tabular}
}
\label{table:igr}
\end{table}


Table \ref{table:igr} shows the results of the experiments with our approach. The bolded values on this table indicate fairness improvements compared to the original settings. 
Despite the inability of DetGreedy and DetRelaxed to enhance fairness in the following scenarios, our approach significantly increased fairness by a greater margin (approximately 24\% on average in all scenarios): four more scenarios in the GetSentry project (CORMS-Top-4, CORMS-Top-6, CORMS-Top-10, RevFinder-Top-10), three more scenarios in the Node.js project (CORMS-Top-4, RevFinder-Top-6, RevFinder-Top-10), and two more scenarios in the Shopify project (CORMS-Top-4, CORMS-Top-6). 
Additionally, while enhancing fairness, the average top-K accuracy was reduced by 1.3\%, 2.6\%, and 2\% in the GetSentry, Node.js, and Shopify projects, respectively, across all scenarios. 

To verify the significance of our approach's fairness improvement compared to the two baselines, i.e., DetGreedy and DetRelaxed, we employ the Wilcoxon signed-rank test \cite{wilcoxon} to test our null hypothesis. The null hypothesis being tested is whether there is a significant difference between the outcomes of our approach's improvement and those of the other two methods. If the p-value is less than 0.05, it means that the two approaches that were compared have significantly different results. The outcomes of our tests on three projects (GetSentry, Node.js, and Shopify), where fairness was enhanced, demonstrate that the p-values for our approach compared to DetGreedy and DetRelaxed are 0.000007 and 0.0002, respectively.
The results imply that our approach's improvements are statistically significant and not due to chance. In terms of fairness evaluation, our approach performs better than both the DetGreedy and DetRelaxed approaches in the examined scenarios.


Note that although our approach results in an overall better performance than both DetGreedy and DetRelaxed, it still cannot fully mitigate the recommendation unfairness on the BSSW project. For example, in this project under RevFinder's top-4 scenario, our approach's $SPD@K$ value is 0.5 (after mitigation), which exceeds the anticipated SPD threshold of 0, which indicates further efforts on improving the fairness for projects with different characteristics are needed. 

\mybox{Our proposed mitigation approach can significantly outperform the existing DetGreedy and DetRelaxed mitigation approaches.}


\section{Threats to Validity}
\label{sec:threats}


The main threat to the validity of this work is the generalizability, as both RevFinder and CORMS use a similarity-based model as a core part of their recommendation systems. 
They attempt to recommend reviewers based on the similarity of the files involved in previous reviews for a given new review request. Additionally, CORMS employs a hybrid approach that combines the results of the similarity model with an SVM model trained on the subjects of previous change reviews. Therefore, it is still possible that the results of experiments conducted on other code reviewer recommendation systems, such as those that use developer experience, could differ from those of CORMS and RevFinder. Consequently, our findings may not be generalizable to those systems. Future research should address these concerns with different code reviewer recommendation systems. 

\section{Conclusion and Future Work}
\label{sec:conclusion}
This paper represents a novel investigation into the fairness issue of machine learning-based code reviewer recommendation systems. Specifically, two state-of-the-art systems (CORMS and RevFinder) and code review data from four open source projects were used to conduct the fairness analysis. 
Our empirical study demonstrates that current state-of-the-art ML-based code reviewer recommendation techniques exhibit unfairness and discriminating behaviors. 
This paper also discusses the reasons why the studied ML-based code reviewer recommendation systems are unfair and provides solutions to mitigate the unfairness. 

In the future, we plan to broaden our research scope by investigating more sensitive attributes, analyzing a wider range of code reviewer recommendation systems with different architectures, and examining fairness issues in other tasks in SE domain.

\balance
\bibliographystyle{IEEEtran}
\bibliography{paper}

\begin{thebibliography}{10}
\providecommand{\url}[1]{#1}
\csname url@samestyle\endcsname
\providecommand{\newblock}{\relax}
\providecommand{\bibinfo}[2]{#2}
\providecommand{\BIBentrySTDinterwordspacing}{\spaceskip=0pt\relax}
\providecommand{\BIBentryALTinterwordstretchfactor}{4}
\providecommand{\BIBentryALTinterwordspacing}{\spaceskip=\fontdimen2\font plus
\BIBentryALTinterwordstretchfactor\fontdimen3\font minus
  \fontdimen4\font\relax}
\providecommand{\BIBforeignlanguage}[2]{{%
\expandafter\ifx\csname l@#1\endcsname\relax
\typeout{** WARNING: IEEEtran.bst: No hyphenation pattern has been}%
\typeout{** loaded for the language `#1'. Using the pattern for}%
\typeout{** the default language instead.}%
\else
\language=\csname l@#1\endcsname
\fi
#2}}
\providecommand{\BIBdecl}{\relax}
\BIBdecl

\bibitem{survey-deep4se}
\BIBentryALTinterwordspacing
Y.~Yang, X.~Xia, D.~Lo, and J.~Grundy, ``A survey on deep learning for software
  engineering,'' \emph{ACM Comput. Surv.}, vol.~54, no. 10s, sep 2022.
  [Online]. Available: \url{https://doi.org/10.1145/3505243}
\BIBentrySTDinterwordspacing

\bibitem{defect-prediction1}
S.~Wang, T.~Liu, and L.~Tan, ``Automatically learning semantic features for
  defect prediction,'' in \emph{Proceedings of the 38th International
  Conference on Software Engineering}, ser. ICSE '16, 2016, p. 297–308.

\bibitem{defect-prediction2}
S.~Wang, T.~Liu, J.~Nam, and L.~Tan, ``Deep semantic feature learning for
  software defect prediction,'' \emph{IEEE Transactions on Software
  Engineering}, vol.~46, no.~12, pp. 1267--1293, 2020.

\bibitem{bug-triage1}
Z.~Li and H.~Zhong, ``Revisiting textual feature of bug-triage approach,'' in
  \emph{2021 36th IEEE/ACM International Conference on Automated Software
  Engineering (ASE)}, 2021, pp. 1183--1185.

\bibitem{WhoReview}
\BIBentryALTinterwordspacing
M.~Chouchen, A.~Ouni, M.~W. Mkaouer, R.~G. Kula, and K.~Inoue, ``Whoreview: A
  multi-objective search-based approach for code reviewers recommendation in
  modern code review,'' \emph{Applied Soft Computing}, vol. 100, p. 106908,
  2021. [Online]. Available:
  \url{https://www.sciencedirect.com/science/article/pii/S1568494620308462}
\BIBentrySTDinterwordspacing

\bibitem{corms}
P.~Pandya and S.~Tiwari, ``Corms: a github and gerrit based hybrid code
  reviewer recommendation approach for modern code review.''\hskip 1em plus
  0.5em minus 0.4em\relax Association for Computing Machinery (ACM), 11 2022,
  pp. 546--557.

\bibitem{Pessach2023}
D.~Pessach and E.~Shmueli, ``A review on fairness in machine learning,''
  \emph{ACM Computing Surveys}, vol.~55, pp. 1--44, 4 2023.

\bibitem{bias-fairness-survey2021}
N.~Mehrabi, F.~Morstatter, N.~Saxena, K.~Lerman, and A.~Galstyan, ``A survey on
  bias and fairness in machine learning,'' \emph{ACM Computing Surveys (CSUR)},
  vol.~54, 7 2021.

\bibitem{making-fair-ml}
\BIBentryALTinterwordspacing
J.~Chakraborty, K.~Peng, and T.~Menzies, ``Making fair ml software using
  trustworthy explanation,'' \emph{Proceedings - 2020 35th IEEE/ACM
  International Conference on Automated Software Engineering, ASE 2020}, pp.
  1229--1233, 9 2020. [Online]. Available:
  \url{https://doi.org/10.1145/3324884.3418932}
\BIBentrySTDinterwordspacing

\bibitem{fairness-enhancing-ml-interventions}
\BIBentryALTinterwordspacing
S.~A. Friedler, C.~Scheidegger, S.~Venkatasubramanian, S.~Choudhary, E.~P.
  Hamilton, and D.~Roth, ``A comparative study of fairness-enhancing
  interventions in machine learning,'' in \emph{Proceedings of the Conference
  on Fairness, Accountability, and Transparency}, ser. FAT* '19.\hskip 1em plus
  0.5em minus 0.4em\relax New York, NY, USA: Association for Computing
  Machinery, 2019, p. 329–338. [Online]. Available:
  \url{https://doi.org/10.1145/3287560.3287589}
\BIBentrySTDinterwordspacing

\bibitem{bias-discrimination-society}
X.~Ferrer, T.~v. Nuenen, J.~M. Such, M.~Coté, and N.~Criado, ``Bias and
  discrimination in ai: A cross-disciplinary perspective,'' \emph{IEEE
  Technology and Society Magazine}, vol.~40, no.~2, pp. 72--80, 2021.

\bibitem{discrimination-aware-dm}
\BIBentryALTinterwordspacing
D.~Pedreshi, S.~Ruggieri, and F.~Turini, ``Discrimination-aware data mining,''
  \emph{Proceedings of the ACM SIGKDD International Conference on Knowledge
  Discovery and Data Mining}, pp. 560--568, 2008. [Online]. Available:
  \url{https://dl.acm.org/doi/10.1145/1401890.1401959}
\BIBentrySTDinterwordspacing

\bibitem{homework-raising-awareness-fairness}
M.~Hort and F.~Sarro, ``Did you do your homework? raising awareness on software
  fairness and discrimination,'' \emph{Proceedings - 2021 36th IEEE/ACM
  International Conference on Automated Software Engineering, ASE 2021}, pp.
  1322--1326, 2021.

\bibitem{fairness-criminal-justice}
``Fairness in criminal justice risk assessments: The state of the art,''
  \emph{Sociological Methods and Research}, vol.~50, pp. 3--44, 2 2021.

\bibitem{fairness-credit-scoring}
N.~Kozodoi, J.~Jacob, and S.~Lessmann, ``Fairness in credit scoring:
  Assessment, implementation and profit implications,'' \emph{European Journal
  of Operational Research}, vol. 297, pp. 1083--1094, 3 2022.

\bibitem{ltdd}
\BIBentryALTinterwordspacing
Y.~Li, L.~Meng, L.~Chen, L.~Yu, D.~Wu, Y.~Zhou, and B.~Xu, ``Training data
  debugging for the fairness of machine learning software,'' in
  \emph{Proceedings of the 44th International Conference on Software
  Engineering}, ser. ICSE '22.\hskip 1em plus 0.5em minus 0.4em\relax New York,
  NY, USA: Association for Computing Machinery, 2022, p. 2215–2227. [Online].
  Available: \url{https://doi.org/10.1145/3510003.3510091}
\BIBentrySTDinterwordspacing

\bibitem{fair-preprocessing}
\BIBentryALTinterwordspacing
S.~Biswas and H.~Rajan, ``Fair preprocessing: Towards understanding
  compositional fairness of data transformers in machine learning pipeline,''
  \emph{ESEC/FSE 2021 - Proceedings of the 29th ACM Joint Meeting European
  Software Engineering Conference and Symposium on the Foundations of Software
  Engineering}, vol.~21, pp. 981--993, 8 2021. [Online]. Available:
  \url{https://dl.acm.org/doi/10.1145/3468264.3468536}
\BIBentrySTDinterwordspacing

\bibitem{fairea}
\BIBentryALTinterwordspacing
M.~Hort, J.~M. Zhang, F.~Sarro, and M.~Harman, ``Fairea: A model behaviour
  mutation approach to benchmarking bias mitigation methods,'' \emph{ESEC/FSE
  2021 - Proceedings of the 29th ACM Joint Meeting European Software
  Engineering Conference and Symposium on the Foundations of Software
  Engineering}, pp. 994--1006, 8 2021. [Online]. Available:
  \url{https://dl.acm.org/doi/10.1145/3468264.3468565}
\BIBentrySTDinterwordspacing

\bibitem{fair-smote}
\BIBentryALTinterwordspacing
J.~Chakraborty, S.~Majumder, and T.~Menzies, ``Bias in machine learning
  software: Why? how? what to do?'' \emph{ESEC/FSE 2021 - Proceedings of the
  29th ACM Joint Meeting European Software Engineering Conference and Symposium
  on the Foundations of Software Engineering}, pp. 429--440, 8 2021. [Online].
  Available: \url{https://doi.org/10.1145/3468264.3468537}
\BIBentrySTDinterwordspacing

\bibitem{fairway}
\BIBentryALTinterwordspacing
J.~Chakraborty, S.~Majumder, Z.~Yu, and T.~Menzies, ``Fairway: a way to build
  fair ml software.''\hskip 1em plus 0.5em minus 0.4em\relax ACM, 11 2020, pp.
  654--665. [Online]. Available:
  \url{https://dl.acm.org/doi/10.1145/3368089.3409697}
\BIBentrySTDinterwordspacing

\bibitem{fairness-crowd}
S.~Biswas and H.~Rajan, ``Do the machine learning models on a crowd sourced
  platform exhibit bias? an empirical study on model fairness,'' \emph{ESEC/FSE
  2020 - Proceedings of the 28th ACM Joint Meeting European Software
  Engineering Conference and Symposium on the Foundations of Software
  Engineering}, pp. 642--653, 11 2020.

\bibitem{fairness-aware-cfg}
\BIBentryALTinterwordspacing
S.~Tizpaz-Niari, A.~Kumar, G.~Tan, and A.~Trivedi, ``Fairness-aware
  configuration of machine learning libraries,'' in \emph{Proceedings of the
  44th International Conference on Software Engineering}, ser. ICSE '22.\hskip
  1em plus 0.5em minus 0.4em\relax New York, NY, USA: Association for Computing
  Machinery, 2022, p. 909–920. [Online]. Available:
  \url{https://doi.org/10.1145/3510003.3510202}
\BIBentrySTDinterwordspacing

\bibitem{ignorance-se-fairness}
J.~M. Zhang and M.~Harman, ``'ignorance and prejudice' in software fairness,''
  \emph{Proceedings - International Conference on Software Engineering}, pp.
  1436--1447, 5 2021.

\bibitem{software-fairness-fse18}
\BIBentryALTinterwordspacing
Y.~Brun and A.~Meliou, ``Software fairness,'' ser. ESEC/FSE 2018.\hskip 1em
  plus 0.5em minus 0.4em\relax New York, NY, USA: Association for Computing
  Machinery, 2018, p. 754–759. [Online]. Available:
  \url{https://doi.org/10.1145/3236024.3264838}
\BIBentrySTDinterwordspacing

\bibitem{software-fairness-survey}
\BIBentryALTinterwordspacing
E.~Soremekun, M.~Papadakis, M.~Cordy, and Y.~L. Traon, ``Software fairness: An
  analysis and survey,'' 5 2022. [Online]. Available:
  \url{https://arxiv.org/abs/2205.08809v1 http://arxiv.org/abs/2205.08809}
\BIBentrySTDinterwordspacing

\bibitem{revfinder}
P.~Thongtanunam, C.~Tantithamthavorn, R.~G. Kula, N.~Yoshida, H.~Iida, and
  K.~I. Matsumoto, ``Who should review my code? a file location-based
  code-reviewer recommendation approach for modern code review,'' \emph{2015
  IEEE 22nd International Conference on Software Analysis, Evolution, and
  Reengineering, SANER 2015 - Proceedings}, pp. 141--150, 4 2015.

\bibitem{detgreedy}
\BIBentryALTinterwordspacing
S.~C. Geyik, S.~Ambler, and K.~Kenthapadi, ``Fairness-aware ranking in search
  and recommendation systems with application to linkedin talent search,''
  \emph{Proceedings of the ACM SIGKDD International Conference on Knowledge
  Discovery and Data Mining}, pp. 2221--2231, 7 2019. [Online]. Available:
  \url{https://dl.acm.org/doi/10.1145/3292500.3330691}
\BIBentrySTDinterwordspacing

\bibitem{fairness-recsys-preproc2}
M.~D. Ekstrand, M.~Tian, I.~M. Azpiazu, J.~D. Ekstrand, O.~Anuyah, D.~McNeill,
  and M.~S. Pera, ``All the cool kids, how do they fit in?: Popularity and
  demographic biases in recommender evaluation and effectiveness,'' in
  \emph{Proceedings of the 1st Conference on Fairness, Accountability and
  Transparency}, vol.~81, 2018, pp. 172--186.

\bibitem{fairness-recsys-preproc1}
B.~Rastegarpanah, K.~P. Gummadi, and M.~Crovella, ``Fighting fire with fire:
  Using antidote data to improve polarization and fairness of recommender
  systems,'' ser. WSDM '19, 2019, p. 231–239.

\bibitem{mcr-survey2021}
N.~Davila and I.~Nunes, ``A systematic literature review and taxonomy of modern
  code review,'' \emph{Journal of Systems and Software}, vol. 177, p. 110951, 7
  2021.

\bibitem{first-crr}
G.~Jeong, S.~Kim, T.~Zimmermann, and K.~Yi, ``Improving code review by
  predicting reviewers and acceptance of patches,'' \emph{Research on software
  analysis for error-free computing center Tech-Memo (ROSAEC MEMO 2009-006)},
  pp. 1--18, 2009.

\bibitem{revrec}
A.~Ouni, R.~G. Kula, and K.~Inoue, ``Search-based peer reviewers recommendation
  in modern code review,'' in \emph{2016 IEEE International Conference on
  Software Maintenance and Evolution (ICSME)}, 2016, pp. 367--377.

\bibitem{rstrace}
E.~Sülün, E.~Tüzün, and U.~Doğrusöz, ``Rstrace+: Reviewer suggestion
  using software artifact traceability graphs,'' \emph{Information and Software
  Technology}, vol. 130, p. 106455, 2021.

\bibitem{hybrid-colab}
Z.~Xia, H.~Sun, J.~Jiang, X.~Wang, and X.~Liu, ``A hybrid approach to code
  reviewer recommendation with collaborative filtering,'' in \emph{2017 6th
  International Workshop on Software Mining (SoftwareMining)}, 2017, pp.
  24--31.

\bibitem{context-aware-crr}
A.~Strand, M.~Gunnarson, R.~Britto, and M.~Usman, ``Using a context-aware
  approach to recommend code reviewers: Findings from an industrial case
  study,'' ser. ICSE-SEIP '20, 2020, p. 1–10.

\bibitem{crr-expand}
A.~Chueshev, J.~Lawall, R.~Bendraou, and T.~Ziadi, ``Expanding the number of
  reviewers in open-source projects by recommending appropriate developers,''
  in \emph{2020 IEEE International Conference on Software Maintenance and
  Evolution (ICSME)}, 2020, pp. 499--510.

\bibitem{how-do-fairness-definitions-fare}
\BIBentryALTinterwordspacing
``How do fairness definitions fare? examining public attitudes towards
  algorithmic definitions of fairness,'' \emph{AIES 2019 - Proceedings of the
  2019 AAAI/ACM Conference on AI, Ethics, and Society}, pp. 99--106, 1 2019.
  [Online]. Available: \url{https://dl.acm.org/doi/10.1145/3306618.3314248}
\BIBentrySTDinterwordspacing

\bibitem{trustworthyai23}
D.~Kaur, S.~Uslu, K.~J. Rittichier, and A.~Durresi, ``Trustworthy artificial
  intelligence: A review,'' \emph{ACM Computing Surveys}, vol.~55, 3 2023.

\bibitem{fairness-recsys-survey}
Y.~Wang, W.~Ma, M.~Zhang, Y.~Liu, and S.~Ma, ``A survey on the fairness of
  recommender systems,'' \emph{ACM Transactions on Information Systems},
  vol.~41, pp. 1--43, 7 2023.

\bibitem{bias-recsys-survey}
\BIBentryALTinterwordspacing
J.~Chen, H.~Dong, X.~Wang, F.~Feng, M.~Wang, H.~Dong, X.~Wang, F.~Feng, and
  X.~He, ``Bias and debias in recommender system: A survey and future
  directions,'' \emph{ACM Transactions on Information Systems}, vol.~41, p.~67,
  2 2023. [Online]. Available: \url{https://dl.acm.org/doi/10.1145/3564284}
\BIBentrySTDinterwordspacing

\bibitem{fairness-recsys-inproc1}
A.~Beutel, J.~Chen, T.~Doshi, H.~Qian, L.~Wei, Y.~Wu, L.~Heldt, Z.~Zhao,
  L.~Hong, E.~H. Chi, and C.~Goodrow, ``Fairness in recommendation ranking
  through pairwise comparisons,'' ser. KDD '19, 2019, p. 2212–2220.

\bibitem{fairness-recsys-inproc2}
Y.~Li, H.~Chen, S.~Xu, Y.~Ge, and Y.~Zhang, ``Towards personalized fairness
  based on causal notion,'' ser. SIGIR '21, 2021, p. 1054–1063.

\bibitem{fairness-recsys-reranking1}
M.~Kaya, D.~Bridge, and N.~Tintarev, ``Ensuring fairness in group
  recommendations by rank-sensitive balancing of relevance,'' in
  \emph{Proceedings of the 14th ACM Conference on Recommender Systems}, ser.
  RecSys '20, 2020, p. 101–110.

\bibitem{fairness-recsys-reranking2}
M.~Morik, A.~Singh, J.~Hong, and T.~Joachims, ``Controlling fairness and bias
  in dynamic learning-to-rank,'' in \emph{Proceedings of the 43rd International
  ACM SIGIR Conference on Research and Development in Information Retrieval},
  ser. SIGIR '20, 2020, p. 429–438.

\bibitem{fairsquare}
\BIBentryALTinterwordspacing
A.~Albarghouthi, L.~D'Antoni, S.~Drews, and A.~V. Nori, ``Fairsquare:
  Probabilistic verification of program fairness,'' \emph{Proc. ACM Program.
  Lang.}, vol.~1, no. OOPSLA, oct 2017. [Online]. Available:
  \url{https://doi.org/10.1145/3133904}
\BIBentrySTDinterwordspacing

\bibitem{fairness-mcr}
\BIBentryALTinterwordspacing
D.~M. German, G.~Robles, G.~Poo-Caama\~{n}o, X.~Yang, H.~Iida, and K.~Inoue,
  ``"was my contribution fairly reviewed?": A framework to study the perception
  of fairness in modern code reviews,'' in \emph{Proceedings of the 40th
  International Conference on Software Engineering}, ser. ICSE '18.\hskip 1em
  plus 0.5em minus 0.4em\relax New York, NY, USA: Association for Computing
  Machinery, 2018, p. 523–534. [Online]. Available:
  \url{https://doi.org/10.1145/3180155.3180217}
\BIBentrySTDinterwordspacing

\bibitem{crr-labeling-bias}
\BIBentryALTinterwordspacing
K.~A. Tecimer, E.~T\"{u}z\"{u}n, H.~Dibeklioglu, and H.~Erdogmus, ``Detection
  and elimination of systematic labeling bias in code reviewer recommendation
  systems,'' in \emph{Evaluation and Assessment in Software Engineering}, ser.
  EASE 2021.\hskip 1em plus 0.5em minus 0.4em\relax New York, NY, USA:
  Association for Computing Machinery, 2021, p. 181–190. [Online]. Available:
  \url{https://doi.org/10.1145/3463274.3463336}
\BIBentrySTDinterwordspacing

\bibitem{genderize}
``Genderize.io: A simple api to predict the gender of a person given their
  name,'' \url{https://genderize.io/}, accessed: 2023-04-13.

\bibitem{fairness-acc-tradeoff2}
Z.~C. Lipton, A.~Chouldechova, and J.~McAuley, ``Does mitigating ml's impact
  disparity require treatment disparity?'' in \emph{Proceedings of the 32nd
  International Conference on Neural Information Processing Systems}, ser.
  NIPS'18.\hskip 1em plus 0.5em minus 0.4em\relax Red Hook, NY, USA: Curran
  Associates Inc., 2018, p. 8136–8146.

\bibitem{fairness-acc-tradeoff1}
S.~Corbett-Davies, E.~Pierson, A.~Feller, S.~Goel, and A.~Huq, ``Algorithmic
  decision making and the cost of fairness,'' in \emph{Proceedings of the 23rd
  ACM SIGKDD International Conference on Knowledge Discovery and Data Mining},
  ser. KDD '17, 2017, p. 797–806.

\bibitem{user-oriented-fairness}
\BIBentryALTinterwordspacing
Y.~Li, H.~Chen, Z.~Fu, Y.~Ge, and Y.~Zhang, ``User-oriented fairness in
  recommendation,'' \emph{The Web Conference 2021 - Proceedings of the World
  Wide Web Conference, WWW 2021}, pp. 624--632, 4 2021. [Online]. Available:
  \url{https://dl.acm.org/doi/10.1145/3442381.3449866}
\BIBentrySTDinterwordspacing

\bibitem{amortizing-individual-fairness}
A.~J. Biega, K.~P. Gummadi, and G.~Weikum, ``Equity of attention: Amortizing
  individual fairness in rankings,'' \emph{41st International ACM SIGIR
  Conference on Research and Development in Information Retrieval, SIGIR 2018},
  vol.~18, pp. 405--414, 2018.

\bibitem{cpfair}
M.~Naghiaei, H.~A. Rahmani, and Y.~Deldjoo, ``Cpfair: Personalized consumer and
  producer fairness re-ranking for recommender systems,'' in \emph{Proceedings
  of the 45th International ACM SIGIR Conference on Research and Development in
  Information Retrieval}, ser. SIGIR '22, 2022, p. 770–779.

\bibitem{fairrec}
G.~K. Patro, A.~Biswas, N.~Ganguly, K.~P. Gummadi, and A.~Chakraborty,
  ``Fairrec: Two-sided fairness for personalized recommendations in two-sided
  platforms,'' in \emph{Proceedings of The Web Conference 2020}, ser. WWW '20,
  2020, p. 1194–1204.

\bibitem{popularity-bias}
Z.~Zhu, Y.~He, X.~Zhao, Y.~Zhang, J.~Wang, and J.~Caverlee,
  ``Popularity-opportunity bias in collaborative filtering,'' in
  \emph{Proceedings of the 14th ACM International Conference on Web Search and
  Data Mining}, ser. WSDM '21.\hskip 1em plus 0.5em minus 0.4em\relax
  Association for Computing Machinery, 2021, p. 85–93.

\bibitem{longtail1}
A.~Ferraro, ``Music cold-start and long-tail recommendation: Bias in deep
  representations,'' ser. RecSys '19, 2019, p. 586–590.

\bibitem{longtail2}
S.~Liu and Y.~Zheng, ``Long-tail session-based recommendation,'' in
  \emph{Proceedings of the 14th ACM Conference on Recommender Systems}, ser.
  RecSys '20, 2020, p. 509–514.

\bibitem{wilcoxon}
R.~F. Woolson, \emph{Wilcoxon Signed-Rank Test}.\hskip 1em plus 0.5em minus
  0.4em\relax John Wiley and Sons, Ltd, 2008, pp. 1--3.

\end{thebibliography}

\end{document}